\documentclass{aa}
\usepackage{amssymb}
\usepackage{amsmath}
\usepackage{graphicx}
\usepackage{color}
\usepackage{txfonts}
\usepackage{epstopdf}
\usepackage{siunitx}
\usepackage{ulem}

\newcommand\Gaia{\textit{Gaia\/}}
\newcommand\Hipparcos{\textit{Hipparcos\/}}
\renewcommand\deg{$^{\circ}$\xspace}

\begin{document} 

\title{ \Gaia\/ Radial Velocity Spectrometer}
\titlerunning{ \Gaia\/ RVS}

\author{M.~Cropper              \inst{\ref{inst:mssl}} 
\and D. Katz                    \inst{\ref{inst:0001}}
\and P. Sartoretti              \inst{\ref{inst:0001}}
\and T.~Prusti                  \inst{\ref{inst:estec}}
\and J.H.J.~de~Bruijne  \inst{\ref{inst:estec}}
\and F. Chassat                 \inst{\ref{inst:airbus}}
\and P. Charvet                 \inst{\ref{inst:airbus}}
\and J. Boyadijan               \inst{\ref{inst:airbus}}
\and M. Perryman                \inst{\ref{inst:estec-ss}}
\and G.~Sarri                           \inst{\ref{inst:estec}}
\and P.~Gare                            \inst{\ref{inst:estec}}
\and M. Erdmann         \inst{\ref{inst:estec}}
\and U.~Munari                  \inst{\ref{inst:asiago}}
\and T. Zwitter                          \inst{\ref{inst:0016}}
\and M. Wilkinson               \inst{\ref{inst:leicester}}
\and F. Arenou                  \inst{\ref{inst:0001}}
\and A. Vallenari               \inst{\ref{inst:padua}}
\and A. G\'{o}mez               \inst{\ref{inst:0001}}
\and P. Panuzzo                     \inst{\ref{inst:0001}} 
\and G. Seabroke                   \inst{\ref{inst:mssl}}
\and C.~Allende~Prieto  \inst{\ref{inst:mssl},\ref{inst:0541}}
\and K. Benson                       \inst{\ref{inst:mssl}}  
\and O. Marchal                     \inst{\ref{inst:0001}}
\and H. Huckle                      \inst{\ref{inst:mssl}}  
\and M. Smith                       \inst{\ref{inst:mssl}}
\and C. Dolding                  \inst{\ref{inst:mssl}}
\and K. Jan{\ss}en                \inst{\ref{inst:0013}}
\and Y. Viala                          \inst{\ref{inst:0001}}
\and R. Blomme                    \inst{\ref{inst:0003}}
\and S.  Baker                     \inst{\ref{inst:mssl}} 
\and S.~Boudreault              \inst{\ref{inst:mssl},\ref{inst:0014}} 
\and F.  Crifo                       \inst{\ref{inst:0001}}  
\and C. Soubiran                 \inst{\ref{inst:0011}}
\and Y.  Fr\'{e}mat                \inst{\ref{inst:0003}}
\and G. Jasniewicz                \inst{\ref{inst:0004}}
\and A. Guerrier                 \inst{\ref{inst:0005}}
\and L.P. Guy                           \inst{\ref{inst:0010}}
\and C. Turon                   \inst{\ref{inst:0001}}
\and A.  Jean-Antoine-Piccolo       \inst{\ref{inst:CNES}}
\and F.  Th\'{e}venin            \inst{\ref{inst:0012}}
\and M. David                         \inst{\ref{inst:0007}}
\and E. Gosset                 \inst{\ref{inst:0008},\ref{inst:0009}}
\and Y.  Damerdji                \inst{\ref{inst:0008},\ref{inst:0006}}
}

\institute {Mullard Space Science Laboratory, University College London, Holmbury St Mary, Dorking, Surrey RH5 6NT, UK                                                           \label{inst:mssl}
\and GEPI, Observatoire de Paris, Universit\'e PSL, CNRS,  5 Place Jules Janssen, F-92190 Meudon, France                                                                                     \label{inst:0001}
\and ESA, European Space Research and Technology Centre (ESTEC), Keplerlaan 1, NL-2201 AG,  Noordwijk, The Netherlands                                                           \label{inst:estec}
\and Airbus Defence and Space, 31 Rue des Cosmonautes, F-31402 Toulouse Cedex France                                                                                                           \label{inst:airbus}
\and Scientific Support Office, Directorate of Science, European Space Research and Technology Centre (ESA/ESTEC), Keplerlaan 1, NL-2201AZ, Noordwijk, The Netherlands \label{inst:estec-ss}
\and INAF-National Institute of Astrophysics, Osservatorio Astronomico di Padova, Osservatorio Astronomico, I-36012 Asiago (VI), Italy                                         \label{inst:asiago}
\and Faculty of Mathematics and Physics, University of Ljubljana, Jadranska ulica 19, SLO-1000 Ljubljana, Slovenia                                                                                          \label{inst:0016}
\and Department of Physics \& Astronomy, University of Leicester, University Road, Leicester, LE1 7RH, UK                                                                                 \label{inst:leicester}
\and INAF - Padova Observatory, Vicolo dell'Osservatorio 5, I-35122 Padova, Italy                                                                                                                           \label{inst:padua}
\and Instituto de Astrof\'{\i}sica de Canarias, E-38205 La Laguna, Tenerife, Islas Canarias, Spain                                                                                                         \label{inst:0541}
\and Leibniz Institute for Astrophysics Potsdam (AIP), An der Sternwarte 16, D-14482 Potsdam, Germany                                                                                          \label{inst:0013}
\and Royal Observatory of Belgium, Ringlaan 3, B-1180 Brussels, Belgium                                                                                                                                                   \label{inst:0003}
\and Max Planck Institut f{\"u}r Sonnensystemforschung, Justus-von-Liebig-Weg 3, D-37077 G{\"o}ttingen, Germany                                                                               \label{inst:0014}
\and Laboratoire d'Astrophysique de Bordeaux, Universit\'{e} de Bordeaux, CNRS, B18N, all{\'e}e Geoffroy Saint-Hilaire, F-33615 Pessac, France                                  \label{inst:0011}
\and Laboratoire Univers et Particules de Montpellier, Universit\'{e} Montpellier, Place Eug\`{e}ne Bataillon, CC72, F-34095 Montpellier Cedex 05, France                        \label{inst:0004}
\and Thales Services, 290 All{\'e}e du Lac, F-31670 Lab{\`e}ge, France                                                                                                                                          \label{inst:0005}
\and Department of Astronomy, University of Geneva, Chemin d'Ecogia 16, CH-1290 Versoix, Switzerland                                                                                             \label{inst:0010}
\and Centre National d'\'Etudes Spatiales, 18 Avenue Edouard Belin, F-31400 Toulouse, France                                                                                                 \label{inst:CNES}
\and Laboratoire Lagrange, Universit\'{e} Nice Sophia-Antipolis, Observatoire de la C\^{o}te d'Azur, CNRS, CS 34229, F-06304 Nice Cedex, France\                          \label{inst:0012}
\and Universiteit Antwerpen, Onderzoeksgroep Toegepaste Wiskunde, Middelheimlaan 1, B-2020 Antwerpen, Belgium                                                                                  \label{inst:0007}
\and Institut d'Astrophysique et de G\'{e}ophysique, Universit\'{e} de Li\`{e}ge, 19c, All\'{e}e du 6 Ao\^{u}t, B-4000 Li\`{e}ge, Belgium                                                 \label{inst:0008}
\and F.R.S.-FNRS, Rue d'Egmont 5, B-1000 Brussels, Belgium                                                                                                                                                                 \label{inst:0009}
\and CRAAG - Centre de Recherche en Astronomie, Astrophysique et G\'{e}ophysique, Route de l'Observatoire, Bp 63 Bouzareah DZ-16340 Algiers, Algeria                       \label{inst:0006}
}

\date{ }

\abstract{This paper presents the specification,  design, and development of the Radial Velocity Spectrometer (RVS) on the European Space Agency's \Gaia\/ mission. Starting with the rationale for the full six dimensions of phase space in the dynamical modelling of the Galaxy, the scientific goals and derived top-level instrument requirements are discussed, leading to a brief description of the initial concepts for the instrument. The main part of the paper is a description of the flight RVS, considering the optical design, the focal plane, the detection and acquisition chain, and the as-built performance drivers and critical technical areas. After presenting the pre-launch performance predictions, the paper concludes with the post-launch developments and mitigation strategies, together with a summary of the in-flight performance at the end of commissioning.}


   \maketitle
%


\section{Introduction}
\label{sec:intro}

The \Gaia\/ satellite of the European Space Agency (ESA) was launched on 2013 December 19, arriving at the L2 point a month later, for a planned five-year mission after the commissioning, which ended in 2014 July (the mission was extended for a further 1.5 years in late 2017). \Gaia\/ was conceived as an astrometric satellite, extending by orders of magnitude in terms of distance and accuracy the pioneering results from ESA's \Hipparcos\/  satellite. The mission, a collaboration between ESA, industrial partners, and science institutes in ESA member states, is described in \cite{Prusti:16}. The first data release was made in 2016 September and is described in \cite{Brown:16}.  

The science return from \Hipparcos\/ is very significant (see \cite{Perryman:09} for a comprehensive overview), but its payload permitted only astrometric and photometric measurements. Measurement of stellar positions over time produced proper (transverse) motions and distances, but not a measure of the velocity in the line of sight (the radial velocity). This was recognised as a deficiency at the time; see for example Blaauw in \cite{Blaauw:88}. \cite{Perryman:09} provides a comprehensive overview of the proposals made in France and the UK and also within ESA in the period 1980 -- 1987 for new dedicated telescopes and updated instrumentation. None of these was successful.  Emphasising that for many studies it is of the greatest importance to have all three space velocities rather than only the two components of the star's velocity on the sky, \cite{Binney:97} noted how few stars in the \Hipparcos\/ Input Catalog \citep{Turon:92} had radial velocities. A ground-based ESO Large Programme was instigated to provide radial velocities of the $\sim60\,000$ stars in the \Hipparcos\/ Input Catalog with spectral type later than F5, but progress was slow. The situation improved only when \cite{Nordstrom:04} published good measurements for $\sim13\,500$ F and G dwarfs and  \cite{Famaey:05} published radial velocities for 5\,952 K and 739 M giants, but this was still a small fraction of the total catalogue\footnote{Since that time, several large spectroscopic surveys for galactic science have been undertaken, including RAVE \citep{Steinmetz:16}, APOGEE \citep{Majewski:17}, ESO-Gaia \citep{Gilmore:12}, LAMOST \citep{cui:12}, and GALAH \citep{Martell:17}.}. This shortcoming was therefore fully evident at the time when the early \Gaia\/ concepts were being developed, and hence a spectrometer, the Radial Velocity Spectrometer (RVS), was incorporated into the payload to avoid such a science loss \citep{Favata:95, Favata:97}. 

Beyond the radial velocities, this initiative also for the first time enabled a spectroscopic survey of the entire sky to measure astrophysical parameters of point sources. \citet{Perryman:95} emphasised the scientific utility from acquisition of information complementary to the astrometric measurements in the \Gaia\/ \citep{Lindegren:96} and {\it Roemer} \citep{Hoeg:94} concepts being developed at that time. In addition to providing full space motions, he identified the advantages to the mission of multiple visits for identifying binary systems, and correction of perspective accelerations, and also the wider benefits of the large-scale determination of elemental chemical abundances that would inform the star formation history and provide chemical enrichment information, to parallel that from the kinematic measurements. The paper highlighted the scale of the task, given the significant increase in kinematic data in the mission concepts.

While \cite{Perryman:95} mainly considered ground-based solutions using multi-fibre spectrographs, it was clear that a dedicated instrument in orbit would provide more complete and uniform complementary information. The initial concept presented in \cite{Favata:95, Favata:97} was a slitless scanning spectrograph called the Absolute Radial Velocities Instrument (ARVI). This would provide a radial velocity precision of $\sim10$ km s$^{-1}$ at a limiting magnitude of $\sim17$, with $\sim 1$ km s$^{-1}$ for brighter magnitudes $10-12$, to achieve a metallicity determination precision of $\sim0.1$ dex. The instrument would use a separate optical system to that for the astrometry. Many of the critical aspects important in the long-term for the RVS were discussed in the ARVI papers, including the limiting magnitude, resolution, bandpass, scanning rate, telemetry- and attitude-control requirements, and the wavelength zero point.

Taken together with the photometric measurements planned in the \Gaia\/ mission concept (to provide luminosities and temperatures, as well as photometric distances and ages), the change in emphasis for this next generation of mission should not be underestimated. Through the acquisition of both kinematic and astrophysical data,  \Gaia\/ was developed from an advanced astrometric satellite into a complete facility for the study of the formation and evolution of the Galaxy.

This paper provides a brief overview of the RVS concept and requirements (Sec.~\ref{sec:concept} and \ref{sec:requirements}) and then describes the instrument (Sec.~\ref{sec:optical_design} to \ref{sec:d&a}) before examining the pre-launch performance (Sec.~\ref{sec:pp}). Post-launch developments, optimisations, and updated performance predictions are summarised in Sec.~\ref{sec:post-launch} --\ref{sec:performance}. 

We distinguish in this paper between pre-launch instrument parameters -- for example as implemented at the critical design review (CDR) or derived from the ground-based calibrations -- and those post-launch, after which they may have been optimised for the in-orbit characteristics of the satellite during commissioning. It should be kept in mind that the in-orbit performance described from Sec.~\ref{sec:post-launch} onwards supersedes the pre-launch expectations, and also that the full end-to-end performance of the RVS instrument is achieved in conjunction with the full  \Gaia\/ data-processing system, described in \cite{Sartoretti:18} and \cite{Katz:18}.

\section{Early RVS concepts}
\label{sec:early}

Subsequent to \Gaia's adoption in 2000 October initiating the major industrial activities, ESA in mid-2001 instigated working groups for the scientific community to contribute to the development of the mission. One of these was the RVS Working Group. In the period $2001-2006$, this group examined in detail the scientific requirements of the instrument. 

Fundamental considerations included the wavelength range of the spectrometer, spectral resolution, and the limiting magnitude. Because \Gaia\/ would operate in time-delay integration (TDI) mode, in which the spectra would scan over the focal plane at the same rate at which the CCD detectors were being read out, RVS would necessarily be slitless. To minimise the background light, the wavelength range should be as narrow as possible, consistent with it containing sufficient strong spectral lines to provide radial velocity information, as well as an adequate range of chemical elements to provide astrophysical information (temperatures, gravities, and metallicities). Spectral regions around the \ion{Mg}{ii} doublet at $\sim440$ nm and the \ion{Ca}{ii} triplet at $\sim850$ nm were examined, with a $25$ nm region centred on the \ion{Ca}{ii} triplet preferred (this spectral domain was originally proposed by U. Munari, and noted independently by R. Le Poole). 

The balance between the radial velocity and astrophysical information (requiring higher signal-to-noise ratios) for the setting of the resolving power requirements between $R=5000$ and $20000$ was explored, with $R=11500$ found to be optimal in providing both adequate spectral resolution while maximising the radial velocity performance, and taking into account other factors as
well, such as the telemetry budget. 

Because the RVS bandpass is narrow compared to that of the astrometric instrument, so that fewer photons are recorded, and because the spectral dispersion distributes these over a larger number of pixels, each of which has associated noise sources, the limiting magnitude would necessarily be lower. It was therefore important to consider the scientific drivers carefully and match the radial velocity accuracy with that of the transverse velocities for the scientific scenarios of interest. These considerations set requirements of $3-15$ km s$^{-1}$ for $~V=16.5$  K-type giants \citep{Wilkinson:05}. These and other requirements were consolidated for the spectroscopic requirements in the Mission Requirements Document for the Implementation Phase of the programme \citep{ESA:05}. 

At the start of the implementation phase, the working groups were shut down and the expertise was transferred to the \Gaia\/ Data Processing and Analysis Consortium (DPAC).

Over the same period, ESA and some European agencies funded an engineering study led by a consortium constituted from science institutes, with the aim of complementing the work of the industrial teams, who were concentrating mainly on the astrometric performance of \Gaia\/ during this competitive tendering phase. Based on the earlier Phase A activities, the payload concept at that point contained a separate telescope (Spectro) for the RVS and medium-band photometry \citep[see for example][]{MMS:99, merat:99, Perryman:01, safa:04}. In this complementing study by the RVS Consortium, the driving performance considerations for the instrument design were to maximise the radial velocity precision, and to minimise the constraints imposed by the telemetry limitations. 

The signal levels were increased by maximising the field of view and reducing the focal ratio consistent with optical distortion and spectral resolution, in order to reduce the scanning speed over the detector and maximise the exposure duration. The principal noise source was identified as that arising in the detector from the readout (readout noise), so electron-multiplying CCDs (also known as L3CCDs) were specified to reduce this to a minimum. Although a generous fraction of the overall \Gaia\/ telemetry was allocated to the RVS, the length of the spectra imposed high data rates, and this, with the desirability of two-dimensional information to separate overlapping spectra optimally, led to a scheme in which data from the CCDs were combined on board in order to remain within the budget. Performance predictions and system margins were within budget.

The work during this period was reported in \cite{Katz:03}, \cite{Munari:03}, \cite{Cropper:03}, \cite{Katz:05}, \cite{Cropper:05a}, \cite{Cropper:05b}, and especially in \cite{Katz:04} and \cite{Wilkinson:05}. 

For the implementation phase in 2006, the selected prime contractor Astrium (now Airbus Defence and Space) proposed and implemented a different RVS instrument concept, in part using ideas in \cite{Cropper:01}.

\section{Flight instrument concept}
\label{sec:concept}

\begin{figure}[h!]
\includegraphics[width=\columnwidth] {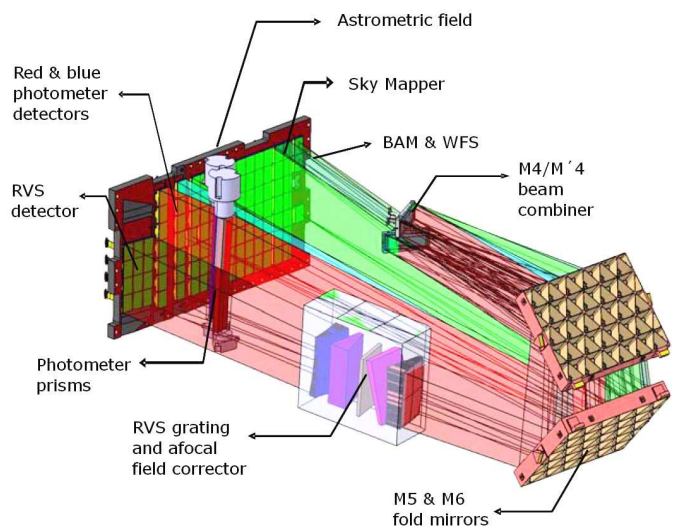}
\caption{Layout of the optical beams after the beam combiner from the two telescopes, and the focal plane in \Gaia\/. The scan direction is from right to left. BAM is the Basic Angle Monitor, and WFS is the Wavefront Sensor.}
\label{fig:concept}
\end{figure}

The flight RVS design departed from the earlier concepts discussed briefly in Sec.~\ref{sec:early} above by removing the Spectro telescope, and employing, instead, the telescopes for the astrometric instrument. This was motivated by savings in mass, power (and heat dissipation), improved payload module accommodation, and cost. The starlight is dispersed by a block of RVS optics, which produces a spectrum that is approximately confocal with the undispersed beams, and with the same focal ratio. The optics block also defines the instrument bandpass and corrects the off-axis characteristics of the beam. The RVS focal plane is located in the same focal plane array as the astrometric instrument. Starlight enters the spectrometer after the astrometric (and photometric) instruments during normal operations when the satellite is scanning. There are 12 CCDs in the RVS focal plane. In order to limit the size of the elements in the optics block, only four rows of CCDs are employed, instead of seven in the astrometric focal plane. The instrument uses the SkyMapper information from the astrometric field of view. The median optical spectral resolving power of 10\,400 was compliant with the nominal requirement \citep{ESA:05}, and with a sampling of $\sim$3 pixels per resolution element, the window was 1260 pixels long. This layout is shown schematically in Fig.~\ref{fig:concept}. 

The RVS is considered to consist of the RVS optics block, the focal plane, and the dedicated software to place windows on the focal plane. The instrument and its operation with other payload elements including the SkyMappers is described briefly in \cite{Prusti:16}. 

The flight design therefore benefits from the larger light grasp of the telescopes feeding the astrometric instrument, and with two telescopes, the doubling of the number of observations of each object, as well as from the removal of an entire optical system with its separate star trackers. However, the integration time per CCD is limited to the same as that of the astrometric instrument, 4.4 s, which significantly reduces the exposure levels with respect to the earlier concept (in which the focal ratio was shorter and the field of view larger), and the smaller number of rows of CCDs in the RVS focal plane reduces the number of observations per object. With respect to the performance expected in the earlier concepts discussed in \cite{Katz:04}, and from a different perspective, in \cite{Cropper:04, Cropper:05b}, the projected pre-launch limiting magnitude of the flight design was $\sim1$ magnitude poorer owing to the shorter exposure times from the telescopes, and conventional CCDs, rather than L3CCDs in the focal plane, with implications for the science case discussed in \cite{Wilkinson:05}. On the other hand, with the experience of processing in-orbit RVS data, the longer focal length arising from the use of the same telescope as that for the astrometric instrument significantly reduces the spectral overlapping, enhancing the performance when one or both telescopes scan crowded regions of the sky.

\section{Requirements}
\label{sec:requirements}

Before describing the instrument in more detail, we identify in Tab.~\ref{tab:requirements} the RVS-specific top level requirements guiding its design. These are extracted from the ESA Mission Requirements Document \citep{ESA:10}, revised to take into account the developments in Sec.~\ref{sec:concept} and hence applicable to the as-implemented instrument. The methodology to be applied to the radial velocity predictions was specified in \cite{deBruijne:05b}.

\begin{table*}[h!]
\begin{center}
\setlength{\tabcolsep}{2pt}
\caption{\label{tab:requirements}
Top-level RVS-specific requirements. Additional requirements include the capability for operation in HR and LR mode (see text), at least Nyquist spectral sampling for HR mode, control over flux rejection levels outside the RVS bandpass, and straylight requirements applicable to the payload as a whole. Spectral types follow the standard terminology, so that temperature decreases from B to K stars, V in the spectral identifier denotes dwarfs, and IIIMP denotes metal poor giants. From \cite{ESA:10}.}
\begin{tabular}{llrll}
\hline \\[-2.5mm]
Average number of transits over mission         & $40$ \hspace{1mm}  for objects  & $V$ & $<$ & $15$                \\
Wavelength range                                        & \multicolumn{4}{l}{$847 - 874$ nm}                              \\
Spectral resolving power        (HR mode)       & \multicolumn{4}{l}{average $10\,500-12\,500$; 90\% $\geq$ 10\,000; max 13\,500}  \\
Spatial resolution                                      & \multicolumn{4}{l}{$1.8$ arcsec to include $90$\% of flux}                               \\
Maximum stellar density                                 & \multicolumn{4}{l}{$36\,000$ objects degree$^{-2}$}          \\
Maximum apparent brightness                     & all spectral types & $V$ & $ <$ & $6$                                        \\
Minimum apparent brightness                     &  B1V & $V$ & $ > $ & $13$                                  \\
                                                                &  G2V & $V $ & $ >$ & $17$                                       \\
                                                                &  K1IIIMP & $V $ & $ >$ & $19$                           \\
Radial velocity systematic error after calibration \hspace*{3mm} &  \multicolumn{4}{l}{$\leq300$ m s$^{-1}$ at end of mission} \\
Radial velocity precision of 1 km s$^{-1}$      &  B1V & $V$ & $\leq$ & $7$          \\
                                                                &  G2V  & $V$ & $\leq$ & $13$             \\
                                                                &  K1III\,MP  & $V$ & $\leq$ & $13.5$             \\
Radial velocity precision of 15 km s$^{-1}$     &  B1V & $V$ & $\leq$ & $12$         \\
                                                                        &  G2V & $V$ & $\leq$ & $16.5$            \\
                                                                        &  K1III\,MP & $V$ & $\leq$ & $17$                \\
\hline
\end{tabular}
\end{center}
\end{table*}

The context for some of these requirements is elaborated in the following subsections. Because of higher scattered light levels encountered in orbit (Sec.~\ref{sec:post-launch}), some of the considerations discussed below required reassessment during the commissioning phase, as described in Sec.~\ref{sec:mitigations} and \ref{sec:performance}.

\subsection{Limiting magnitude and wavelength range}
\label{sec:requirements_lim-mag}

The RVS is an atypical spectrometer in that it is slitless while providing medium resolving power, with constrained exposure durations. Consequently, at intermediate and faint magnitudes, it is photon starved and noise dominated. For its role in providing radial velocities, information from the entire spectrum is condensed into a single velocity value through a cross-correlation, with the radial velocity signature at the faint end emerging only after adding many transits of the object during the survey. Even at the end of the mission, the spectra of most stars will be noise dominated, and will produce only a radial velocity. Simulations  \citep{Katz:04} showed that final signal-to-noise ratios of $\sim$1 per spectral resolution element would nevertheless provide sufficient end-of-mission radial velocity precision. Hence, regardless of the instrument design, at its limiting magnitude and after the noise was minimised, the RVS measurements would have on average $<1$ e$^{-}$ pix$^{-1}$ per exposure. To preserve this signal on average in the presence of noise, it is also essential to provide sufficient levels of digitisation in the detection chain, preferably to $\leq 0.5$ e$^{-}$.

At the faint end, the low fluxes in each pixel render the Poisson noise from the source negligible compared to the other noise sources. Narrowing of the bandpass reduces the cosmic background, which at $\sim850$ nm is mostly zodiacal, but also reduces the kinematic and astrophysical information in each spectrum. Spanning the \ion{Ca}{ii} triplet requires $\sim25$ nm. Loss of one of the triplet lines would substantially reduce the radial velocity signature, while on the other hand, widening the bandpass modestly would not incorporate strong new spectral lines, would increase the sky background, and would also increase the fraction of overlapped spectra. Consequently, the required bandpass was set from 847 to 874 nm. The flux collected in this bandpass is assigned a \Gaia\/ magnitude, $G_{\rm RVS}$ \citep[see][]{Jordi:14}.

The instrumental noise includes components from scattered light and secondary spectral orders, and from the detectors (readout, fixed pattern, and dark noise). Suppression of secondary spectral orders places requirements on the rejection levels for wavelengths outside of the bandpass, which drives the optical coating technology in the instrument. Scattered light may arise within the instrument itself or within the telescope and payload module. Before launch, the noise from the scattered light background was not expected to be a dominant effect, but this was found not to be the case post-launch. This is discussed in Sec.~\ref{sec:post-launch}. 

In respect of detector noise, the signal at any wavelength is accumulated over a full column in the CCD in
TDI operation, so the already small photo-response non-uniformity (PRNU) responsible for the fixed pattern variation requires almost no flat-fielding correction and is negligible. At the $\sim$160K temperatures envisaged for the CCDs in \Gaia,\/ the dark noise is also negligible. In typical astronomical operation with slow readout rates, CCDs and their associated video chains can reach readout noise $\leq5$ e$^{-}$, but as the value must be squared, even a readout noise somewhat below this level will exceed the signal at the faint end by an order of magnitude or more. This is the major noise source in the instrument, and because  it will exceed the cosmic background by a large factor for
almost any instrumental configuration, it is one of the most stringent drivers of performance. 

\subsection{Spectral resolving power}
\label{sec:requirements_resn}

For measuring elemental chemical abundances and surface gravity, the spectral lines should be adequately sampled even in the line cores, and this favours higher spectral resolving power for brighter stars where the signal-to-noise ratio is sufficient for these measurements to be useful. The spectral resolving power sets the length of the spectrum, as higher resolving power produces longer spectra. Longer spectra result in a higher degree of problematic overlapping and a larger telemetry requirement. Given the dominance of readout noise in faint spectra, there was a strong imperative to pack the spectrum into as few pixels as possible, hence lower resolving powers are favoured for the faintest objects; this also reduces the telemetry. To provide optimal performance over the full magnitude range, a requirement was introduced to provide two spectral resolving powers, high resolution (HR), at the full optical resolution of the spectrograph, and low resolution (LR), a resolution-degraded mode for stars with $G_{\rm RVS}>10$ in which pixels could be summed at the detector readout node to reduce the readout noise per sample. The requirements on the detector readout noise as derived from the limiting magnitude performance was set at $\leq 4.0$ e$^{-}$ for LR mode and $\leq 6.0$ e$^{-}$ for HR.

\Gaia\/ is required to scan the sky as uniformly as possible, and to measure the angular distances between stars in the astrometric fields of view at a range of orientations. The adopted scanning law is one of forced precession \citep{Prusti:16}, which results in a sideways (across-scan, AC) displacement (with a sinusoidal dependence) of the star on the spin period of the satellite. In order to maintain the spectral resolving power, the orientation of the dispersion direction was set along the scanning direction (AL) so that spectral lines were not broadened by this AC displacement \citep{Cropper:01,Katz:04}. The (lesser) consequence was that the spectra were broadened with a spatial distribution at a period half that of the spin period, leading to a variation in signal-to-noise ratio on the same period.
 
\subsection{Radiation damage}

As noted in Sec.~\ref{sec:requirements_lim-mag} above, at its limiting magnitude, the RVS instrument concept will work with $<1$ e$^{-}$ pix $^{-1}$ in each exposure to achieve its end-of-mission radial velocity precision. Preserving single electrons in the large number of transfers in the \Gaia\/ CCDs, especially in TDI operation, was unproven, even without in-orbit radiation damage. Trapping sites caused by in-orbit radiation damage renders  preservation of single electrons even more problematic. Even for brighter objects, electrons released from traps after a characteristic time delay leave trails because in TDI operation, the sky moves on from the pixel in which the electron was trapped. With the orientations discussed in Sec.~\ref{sec:requirements_resn}, the trails distort the shape of the spectral lines, inducing a radial velocity shift and diminishing their contrast. When the charge recorded in a given pixel reaches the readout register, it encounters further trapping as it is clocked to the readout node, and this will distort the spatial profile of the spectrum. The effectiveness of the trapping is non-linearly dependent on the quantity of charge being transferred, and on the quantity of charge in preceding pixels, which will themselves be de-trapping, and hence on the spectral energy distribution (in the TDI direction) and the phase of the AC displacement discussed in Sec.~\ref{sec:requirements_resn}. Each across-scan element in the spectrum will have a different effect, with less impact on the central brightest elements, and more on the fainter wings of the spectrum. Minimising, understanding, and modelling the in-orbit radiation damage was therefore a driving factor in the RVS design, and additionally, in the RVS data processing.

\section{RVS optical design}
\label{sec:optical_design}

This section describes the optical design of the combined flight telescope and the RVS. A full report at the critical design review stage is in \cite{Astrium:10b}. 

\subsection{Optical system}
\label{sec:optical_system}

\begin{figure}[h!]
\begin{center}
\includegraphics[width=0.9\columnwidth] {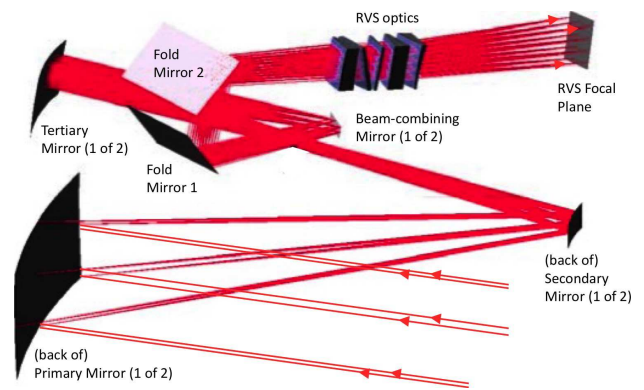}
\end{center}
\caption{RVS optical path for one telescope. The primary mirror is at the lower left, and the RVS focal plane at the upper right, with the RVS optics immediately preceding it in the light path. Below the RVS optics is one half of the beam-combining mirror.}
\label{fig:optical_path}
\end{figure}

\begin{figure}[h!]
\includegraphics[width=\columnwidth] {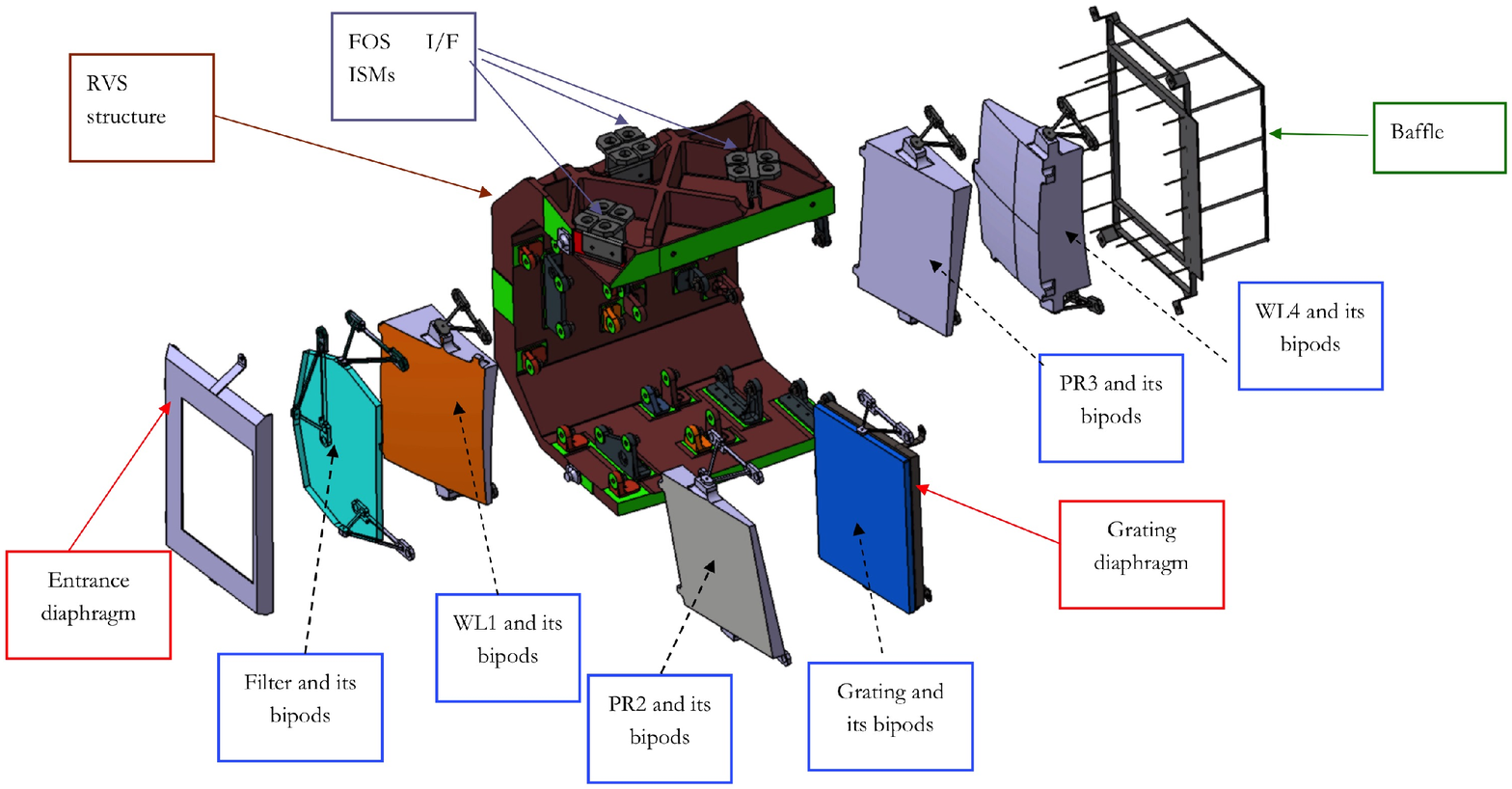}\\[2mm]
\includegraphics[width=\columnwidth] {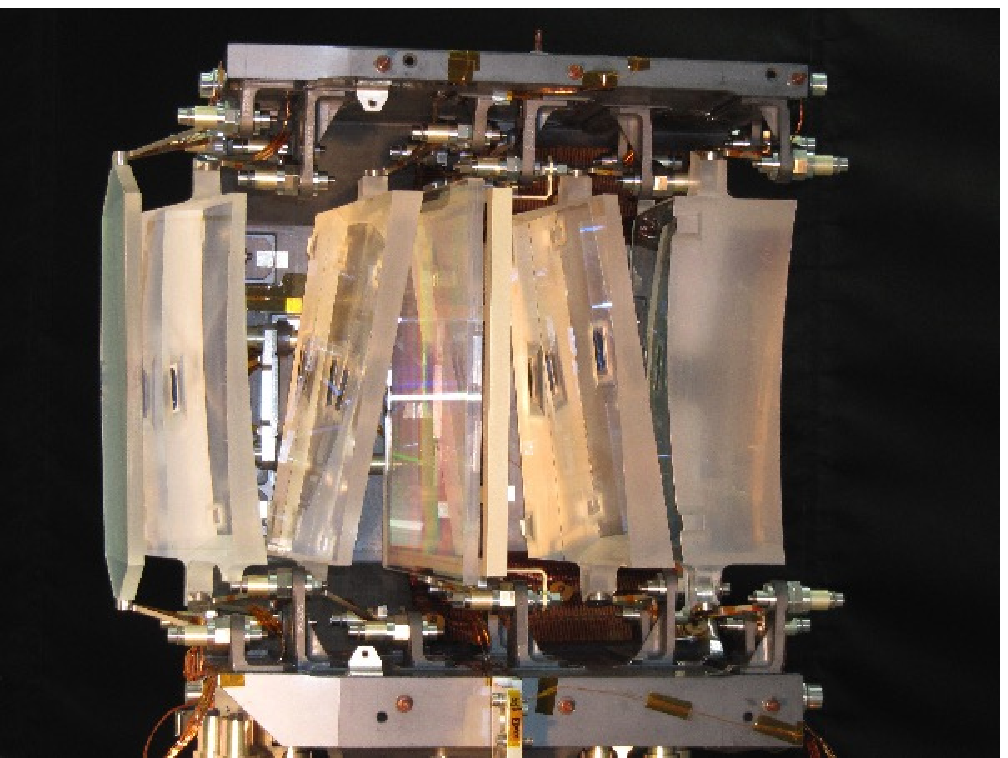}
\caption{\textit {(Top)} Expanded view of the RVS optical module. Light enters from the left, initially encountering the bandpass filter. \textit {(Bottom)} The RVS optics as realised.  The diffraction grating is the fourth element from the left.}
\label{fig:optical_module}
\end{figure}

The two \Gaia\/ telescopes are off-axis three-mirror anastigmats, with rectangular primary mirrors of $1450\times500$ mm$^{2}$ and a focal length of 35\,000 mm. Their beams are combined at the exit pupil by beam-combining mirrors, and folded by two flat mirrors to fit the long focal length into the payload module. They then pass through the RVS optics (Figs.~\ref{fig:concept} -- \ref{fig:grating}) to the RVS detector array ($4\times3$ CCDs) (Fig.~\ref{fig:focal_plane}), whose centre is $\sim0.75$\deg displaced from the optical centre of the focal plane in the along-scan direction, with each telescope producing images displaced by $\pm0.06$\deg in the across-scan direction (in common with the astrometric field and the photometers). The rays on the detectors are not telecentric, and the focal plane is tilted by 7\deg in the focus direction compared to that of the astrometric field. In addition to dispersing the light at moderate spectral resolving power in (slowly) converging beams, the RVS optics must also correct optical aberrations from the off-axis field angles without changing the focal length significantly, and define the instrument bandpass. 

To achieve this, the RVS optics consist of a filter, two plane prisms, two prismatic lenses (\textit{\textup{}} i.e. lenses cut off-axis) and a diffraction grating between the prisms. The arrangement is evident within the optical path in Fig.~\ref{fig:optical_path} and for the module itself in Fig~\ref{fig:optical_module}. The six fused-silica elements are held by thin Invar bipods within a C-section silicon carbide (SiC) structure that provides sufficient rigidity while allowing access for integration. The thermal behaviour of fused-silica is similar to that of SiC, so that the design is almost athermal. Stress relief on the optics is achieved by bonding the bipods to small protrusions on the perimeter of the optical elements. The alignment tolerances for the RVS module within the telescope beams is $0.1-1$ mm and $0.3-1$ milli-radians.

\subsection{Filter}
\label{sec:filter}

The RVS nominal $847-874$ nm bandpass is defined mainly by the multi-layer filter on a fused-silica substrate, modified by the detector quantum efficiency and (slightly) by the telescope transmittance. Attention was paid to the out-of-band rejection in order to minimise the background light in this slitless instrument, especially considering the wide wavelength range over which the detectors are sensitive compared to that of the desired bandpass. Rays from different field-of-view points and pupil points pass through the filter at different locations and at different angles, so that there is a slightly different bandpass for each CCD and for the two telescopes. 

\subsection{Grating}
\label{sec:grating}

\begin{figure}[h!]
\begin{center}
\includegraphics[width=0.501\columnwidth] {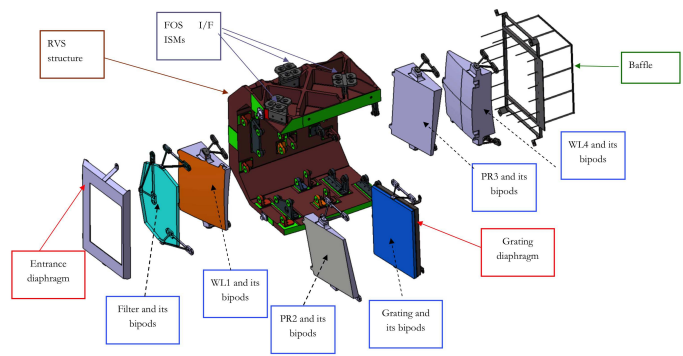}
\includegraphics[width=0.49\columnwidth] {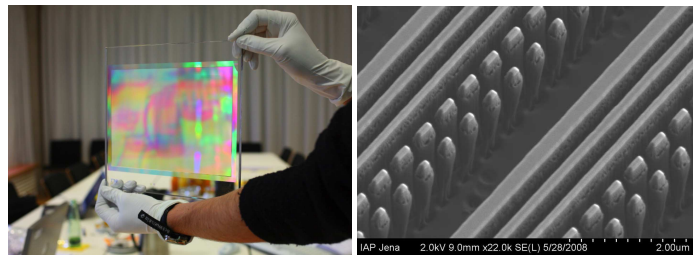}
\end{center}
\caption{(left) Full-scale grating demonstrator. (right) A scanning electron micrograph of the binary elements in the grating used to produce the modulation pattern. Two repeats of line and column elements of decreasing width are shown here.}
\label{fig:grating}
\end{figure}

The RVS grating \citep{Erdmann:10} is an advanced element using the binary index modulation principle to meet the requirements to work in first order with a high efficiency ($\geq 70$\%) and low polarisation sensitivity ($\leq 7$\%) while complying with the tight limits on the additional wavefront error it can contribute. In the context of \Gaia, such gratings provide superior efficiency and reduced unwanted orders. Instead of triangular line profiles produced by ruling the substrate (as in conventional gratings) or variations in refractive index (as in volume phase holographic gratings), subwavelength scale repeats of variable width lines and columns (all of which are at the same height) are used to approximate the grating pattern (Fig.~\ref{fig:grating}). These can be ion-etched onto a fused-silica substrate and allow a high level of control of the grating properties.

\subsection{Baffles}
\label{sec:baffles}

Scattered light can arise from solar system sources, bright stars, and the Galaxy as a whole, and from internal sources. To shadow the telescope apertures from the Sun (particularly), Earth, and Moon (the principal external sources), the \Gaia\/ sunshield is a double-layered circular structure with a diameter of 10.2m. The second layer is slightly smaller so that it intercepts the diffracted flux from the perimeter of the sunward layer.  Within the payload module, the arrangement of the folded telescope beams limits the opportunities for baffling and requires unconventional techniques. Scattered light at the RVS optical module is suppressed by diaphragms, one at the entrance and another associated with the grating, and by an exit baffle (Fig.~\ref{fig:optical_module}). This baffle, and further baffles on the focal plane structure, also serve to prevent light from the RVS (especially unwanted orders) falling on the astrometric and photometric focal planes (Fig.~\ref{fig:focal_plane}). Solar system and stellar sources not in the shadow of the sunshield are not fully baffled.

The main source of internally generated scattered light is the basic angle monitor (BAM), which measures the angle between the two telescopes. This uses laser light at 850nm (i.e. within the RVS bandpass) chosen on the grounds of laser power and stability. Some of the major scattering paths from this source cannot be baffled and were partially mitigated by neutral density filtering at the lasers.

\section{Focal plane}
\label{sec:focal_plane}

\begin{figure}[h!]
\includegraphics[width=\columnwidth] {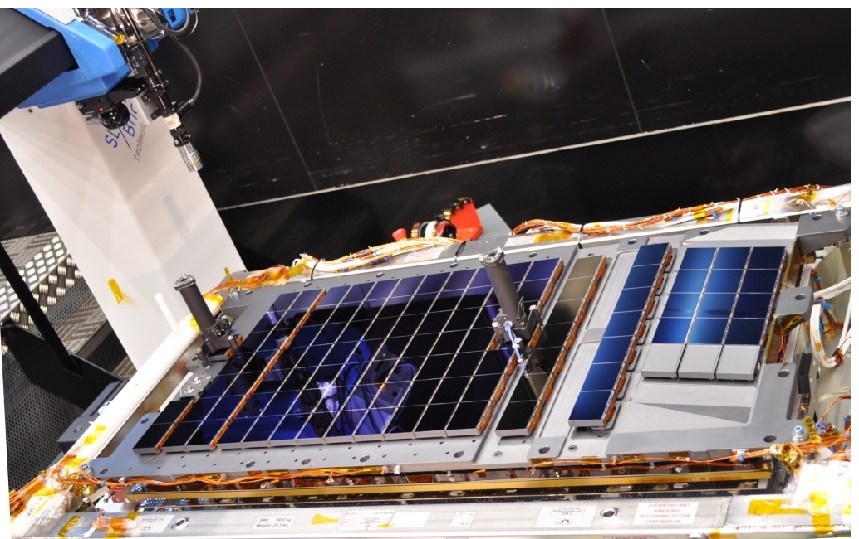}
\includegraphics[width=\columnwidth] {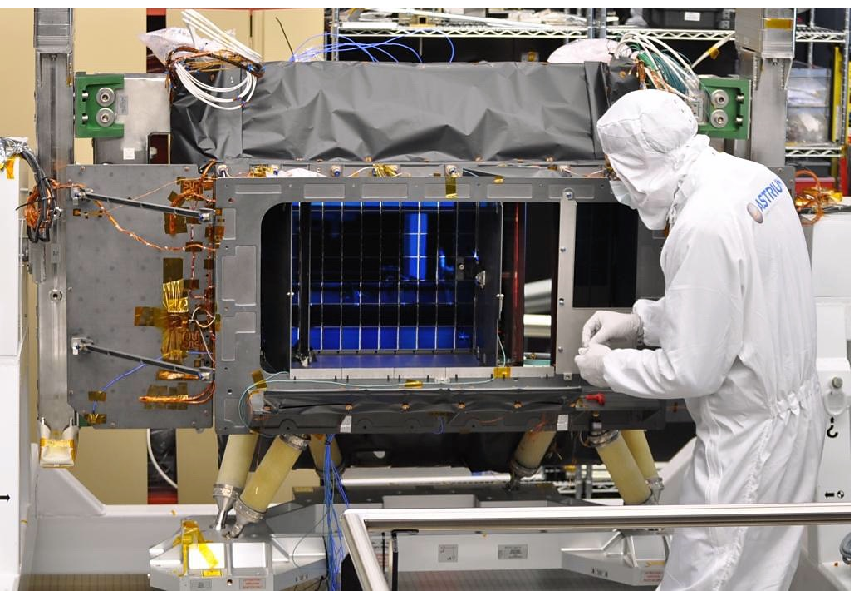}
\caption{\textit {(Top)}  \Gaia\/ focal plane during assembly. The 14 SkyMapper CCDs are the first two full vertical strips on the left of the SiC structure and the 12 RVS CCDs on the right (the fifth row of white rectangles are not CCDs but are used in thermal control). \textit {(Bottom)} The integrated focal plane showing the baffles and radiator (extended structure on each side). Again, the RVS CCDs are on the right.}
\label{fig:focal_plane}
\end{figure}

\begin{figure}[h!]
\includegraphics[width=\columnwidth] {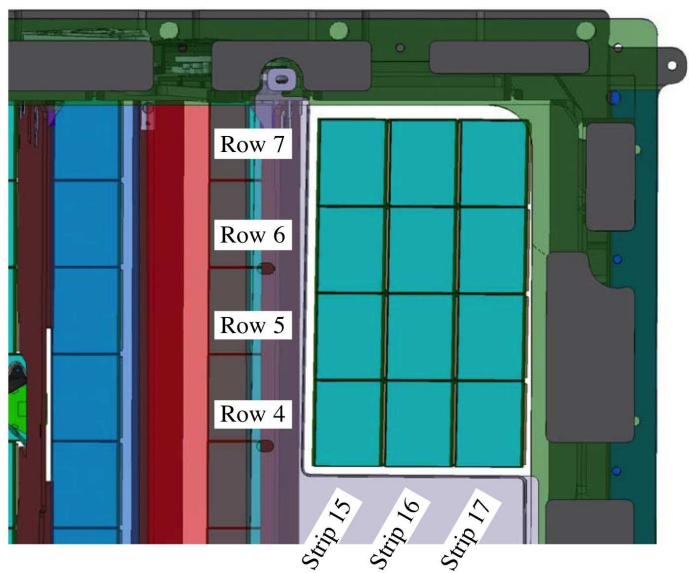}
\caption{Nomenclature for the CCDs in the RVS focal plane.}
\label{fig:RVS_CCD_nomenclature}
\end{figure}

The \Gaia\/ focal plane, Fig.~\ref{fig:focal_plane} \citep{Prusti:16,Crowley:16}, consists of 106 CCDs, distributed across astrometric, photometric, and RVS instruments, as well as SkyMappers, wavefront sensors, and the BAM detectors, all supported on a SiC structure with baffles and thermal control. The RVS focal plane of four rows and three strips of CCDs is the last to be reached in the scan direction, occupying Rows $4-7$, Strips $15-17$ (Fig.~\ref{fig:RVS_CCD_nomenclature}). It is slightly  displaced in the across-scan direction with respect to the rows in the astrometric focal plane. Each CCD has an associated proximity electronics module (PEM) behind it, which in turn interfaces to an interconnection module and a video processing unit (VPU) on each row that service the detectors of all instruments on the row. The RVS PEMs are the same as those for the other instruments, but operate in RVS-specific modes to support the RVS operation. The detector structure is passively cooled by the payload module environment, and is isolated from an ambient-temperature structure holding the PEMs and interconnection module by thermal screens. The heat from the warm part of the focal plane is ejected to space by a radiator.

The detectors in the focal plane detector array are  e2v Technologies CCD91-72, each with $4500\times1966$ rectangular pixels of dimension $10\times30$ $\mu$m along- and across-scan. In order to maximise the sensitivity at the wavelengths of the RVS band, the 40 $\mu$m deep-depletion variant with red-enhanced anti-reflective coating is mounted in the RVS focal plane (it is also used in the Red Photometer and the BAM). In pixel structure and readout node, all \Gaia\/ CCD91-72 variants are the same. The RVS CCDs therefore include charge-injection lines that permit an electrical injection of charge into lines in the image area. These are not used in standard RVS observational sequences as the lines interfere too commonly with the long spectral windows (see Sec.~\ref{sec:d&a}) ($\sim100\times$ longer than the astrometric field windows). The CCD91-72 pixel structure contains a supplementary buried channel that is used to enhance the charge transfer efficiency of electrons during the TDI at low flux levels; these electrons are particularly important in the RVS case. It is not clear \citep{Seabroke:13} whether these structures were correctly implemented in the CCDs selected for flight. 

As (at the time) the readout noise was expected to be the dominant noise source, a very significant effort was undertaken to minimise this in the detection chain. Because of the sub-e$^{-}$ signal levels in the majority of RVS pixels, a fine level of digitisation in the PEM was adopted, $\sim0.55$ e$^-$ per digital unit, which also has the benefit of reducing the digitisation noise significantly. A consequence is that very bright spectra saturate at the level of the analog-digital converter before saturation levels are reached on the CCDs themselves. Consequently, while available, the gate structure within the CCDs  enabling greater dynamic range in the astrometric field and the photometers is not generally used in the RVS, except for some calibrations.

\section{Detection and acquisition chain}
\label{sec:d&a}

\subsection{Window scheme}
\label{subsection:window_scheme}


Given the telemetry bandwidth, the large focal plane, and the short effective exposure duration for all CCDs, it is not possible to transmit to ground all of the sky data. Together with information from the Red Photometer (Fig.~\ref{fig:concept}), stars within the RVS magnitude range are selected, and pixels containing their spectral information are identified by the VPUs to set a window mode in the CCD readout. Pixels within these windows are read out normally and stored in the spacecraft mass memory. Other pixels are generally discarded, except those within virtual object (VO) windows (described below). 

To accommodate the 25 nm of spectrum, the window in the spectral direction was set at 1260 pixels (12.6 mm) long on the CCD. During readout, the selected pixels can be summed on the detector by 3 in the spectral direction to reduce the telemetry, The un-summed and summed modes are termed high resolution (HR) and low resolution (LR)\footnote{This has been modified post-commissioning so that almost all data have been taken in HR; see Sec.~\ref{sec:mitigations}.} , respectively. In addition, to reduce the telemetry further, pixels are summed on the detector in the spatial direction to produce one-dimensional spectra for stars below a certain RVS flux limit. Nominally, the width of the window in the spatial dimension is 10 pixels (300$\mu$m). The windowing is then termed Class 0 for two-dimensional windows containing HR spectra of bright stars for $G_{\rm RVS} \leq7$, Class 1 for one-dimensional windows containing HR spectra for stars $7<G_{\rm RVS} \leq 10$, and Class 2 for one-dimensional LR spectra fainter than $G_{\rm RVS} = 10$. Because whole lines are transferred into the readout register during the TDI operation, if any star is sufficiently bright to be observed in HR, the spectra of other fainter stars in the same TDI lines would also be observed in HR even though they might normally be observed in LR. In order to reduce the telemetry to ground, such spectra are summed digitally in the VPU to produce LR spectra (but with the increased readout noise from the readout of three individual lines rather than a single summed line). To complete this organisation, the spectra are divided into 12 subunits in the along-scan direction, called macrosamples, to facilitate the handling of overlapping windows (see Sec.~\ref{subsec:truncation} below).

In addition to the reduction in the telemetry bandwidth associated with this windowing scheme, the summation of pixels at the detector (10 for HR and 30 for LR) avoids a readout noise contribution for every pixel. This is a critical strategy in minimising what was expected to be the principal noise source in the instrument. A further measure associated with the windowing is to limit the number of single or summed pixels, referred to as samples, read in the serial register (the across scan direction) to a value
of 72, which is consistent with being able to meet the maximum source density requirement of 36\,000 sources degree$^{-2}$ (Tab.~\ref{tab:requirements}). As the unwanted pixels are the large majority, they can be flushed at a higher rate, and most of the parallel line transfer period is available to read the desired samples, thus minimising the readout noise. In order to maintain the thermal stability of the detection chain, 72 samples are always read (unused ones in the overscan region) and discarded if empty.

\subsection{Window truncation}
\label{subsec:truncation}

\begin{figure*}[h!]
\begin{center}
\includegraphics[width=1.5\columnwidth] {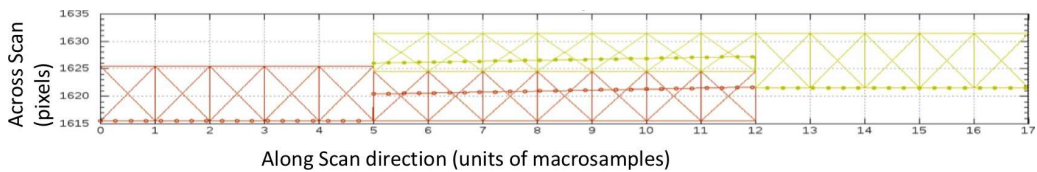}
\end{center}
\caption{Window overlapping in RVS for Class 1 and 2 windows as implemented pre-launch. The macrosamples are numbered along the bottom, so that there are 12 macrosamples per spectrum (yellow and orange). The vertical axis is an example pixel number in the spatial direction. Spectral overlap occurs over macrosamples 6--12, and the window width is apportioned equally (to the nearest integer) over this region. As they are Class 1 and 2 windows, they will be collapsed in the spatial direction at the readout node of the CCDs. In the overlap region, the track of the peak of the spectrum is shown as dots, while outside of it, the dots have no meaning. }
\label{fig:window_overlap}
\end{figure*}

The 1.3 arcmin length of the windows in the spectral direction leads to overlapping of some spectra even in modestly crowded regions (whether spectra overlap also depends on the orientation of the satellite in its scanning, so that for some passes, no overlapping may occur). Because pixels can be selected only once at the readout node, they must be assigned to one or the other of the overlapped spectra. For Class 1 and 2 windows, the overlapping spectra are equally split in the spatial direction between the two windows, and so the effective width to be summed into the one-dimensional spectrum of one of these overlapped windows is between half and the full nominal width, depending on the separation between the sources in the across-scan direction (Fig.~\ref{fig:window_overlap}). Class 0 (i.e.{\it } two-dimensional) windows are assigned at the nominal width and hence take priority over the Class 1 and 2 windows. In the rarer cases of overlapping Class 0 windows, the pixel values are duplicated in each window by the VPU. 

In the spectral direction, overlapped spectra were dealt with in a hierarchy of overlaps with the window positions adjusted on a coarse grid (in the spectral direction) of 105 pixels, the macrosamples mentioned above, so that overlaps can start and end only at macrosample boundaries. (Spectra from different observations are consequently placed slightly differently with respect to the window boundaries.) Unless they are aligned in the scan (spectral) direction, Class 1 and 2 overlapped windows will include some macrosamples of unblended spectra constructed of nominal width, spatial width, and some of blended spectra, constructed of reduced spatial width, as indicated above. Cases of triple and higher order overlaps are treated similarly. In more crowded regions, the spatial width allocated to a source in a one-dimensional window may change several times from macrosample to macrosample. 

Information from the VPU about the window truncation and spectral overlapping is also recorded and telemetered for the data processing.

\subsection{Calibration faint stars}

A subset of stars that would normally be assigned Class 1 and  2 windows because of their magnitude are telemetered with full two-dimensional information in Class 0 windows. These are the calibration faint stars which are used to quantify and calibrate the effects arising from the collapsing on the CCD detectors of the two-dimensional information to one-dimensional in Class 1 and 2 windows for these fainter objects, and in particular, the spatial profile of the spectrum. The information in the spectra from these stars, however, suffers from larger readout noise because pixels are no longer binned before readout.

\subsection{Virtual objects}

While the windows for celestial sources are generated autonomously by the SkyMappers, it is possible to assign windows by command, although they cannot be assigned explicitly to positions on the sky (this can, however, be achieved by timed commands in conjunction with the \Gaia\/ scanning law). These virtual object (VO) windows are used for background monitoring and calibrations, and are of two types: special VOs specify window pattern sequences to be run outside of  observations, for example during orbit maintenance periods; routine VOs are used during routine observations, and these windows are propagated from the SkyMappers to the RVS focal plane position as with normal windows. 

The VO windows can be of all Class 0, 1, or 2, and may accidentally contain sources. The overlap scheme of VOs with other windows follows the standard prioritisation above, so a requested VO of Class 0 may override a normal object window of Class 1.  VO patterns are generated for the focal plane as a whole, but parameters are available specifically for the RVS to enhance their suitability to its characteristics, in particular its larger windows.

\subsection{Data priority scheme} 

Data from the \Gaia\/ instruments are stored in the onboard mass memory unit in file structures that have associated priority levels for telemetry to the ground, and approximately inversely, for data deletion \citep{Astrium:10a}. Brighter objects have higher telemetry priorities, so that in the RVS, windows of Class 0 have the highest and Class 2 the lowest priority. The telemetry priorities of the astrometric, photometric, and spectroscopic data are arranged in an interleaved manner by magnitude. VOs and calibration faint stars have a different telemetry priority, higher than any of the RVS normal object windows, with the VO priority the higher of the two. This prioritisation takes effect when \Gaia\/ is accumulating more data than the mass memory unit can retain, generally when the satellite is scanning along the Galactic Plane. Deleted data are lost, with the consequence that fainter sources may not be recorded in the archive at every transit. This leads to a reduced radial velocity performance at the faint end of the magnitude range.

\section{Pre-launch performance predictions}
\label{sec:pp}

The essential characteristics and performance of \Gaia\/ including the RVS are recorded  for several epochs pre- and post-launch in the \Gaia\/ Parameter Database \citep{deBruijne:05a}. This is the reference repository for the instrument.

\subsection{Instrument throughput and bandpass}
\label{sec:bandpass}

\begin{figure}[h!]
\begin{center}
\includegraphics[width=0.9\columnwidth] {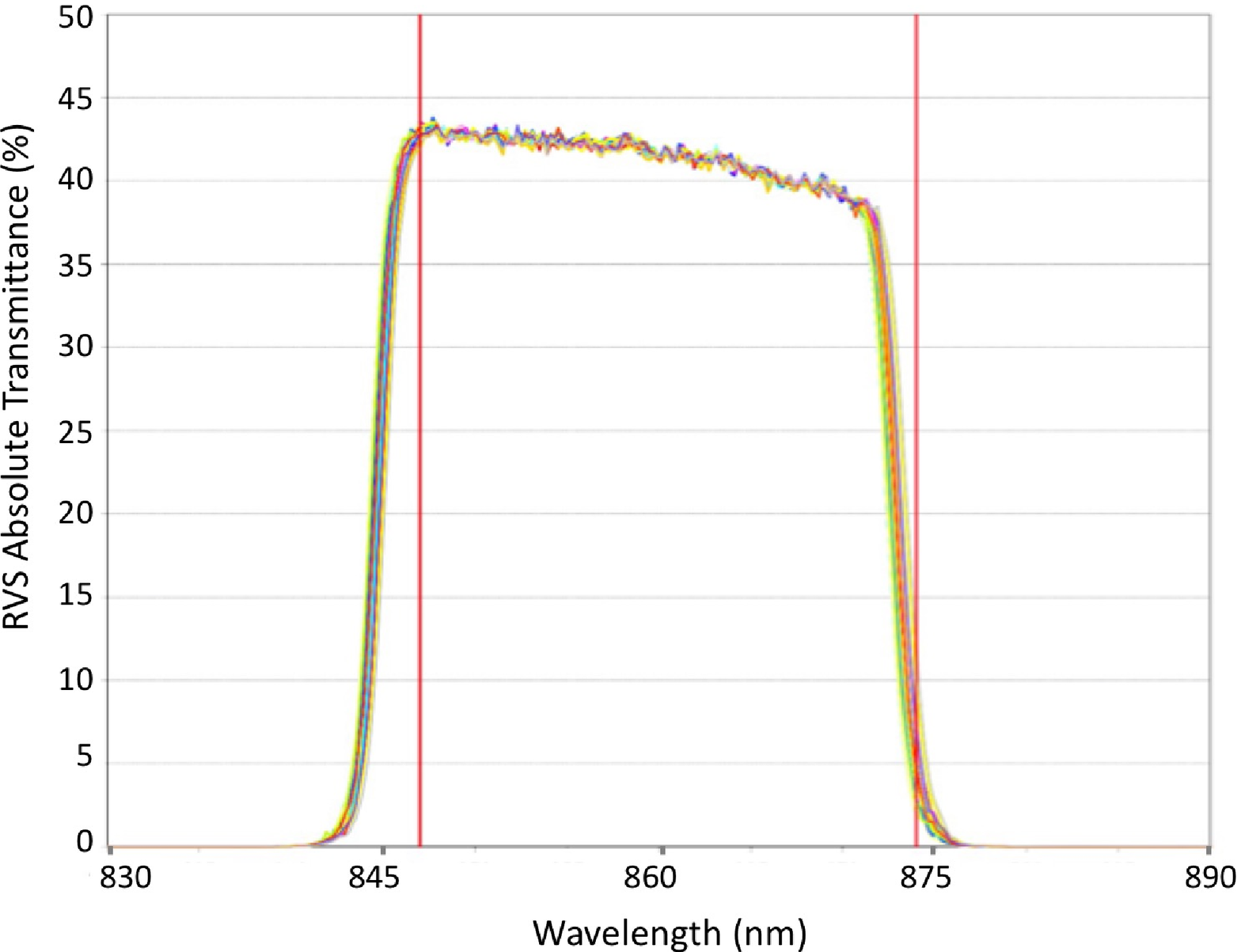}\\[3mm]
\includegraphics[width=0.9\columnwidth] {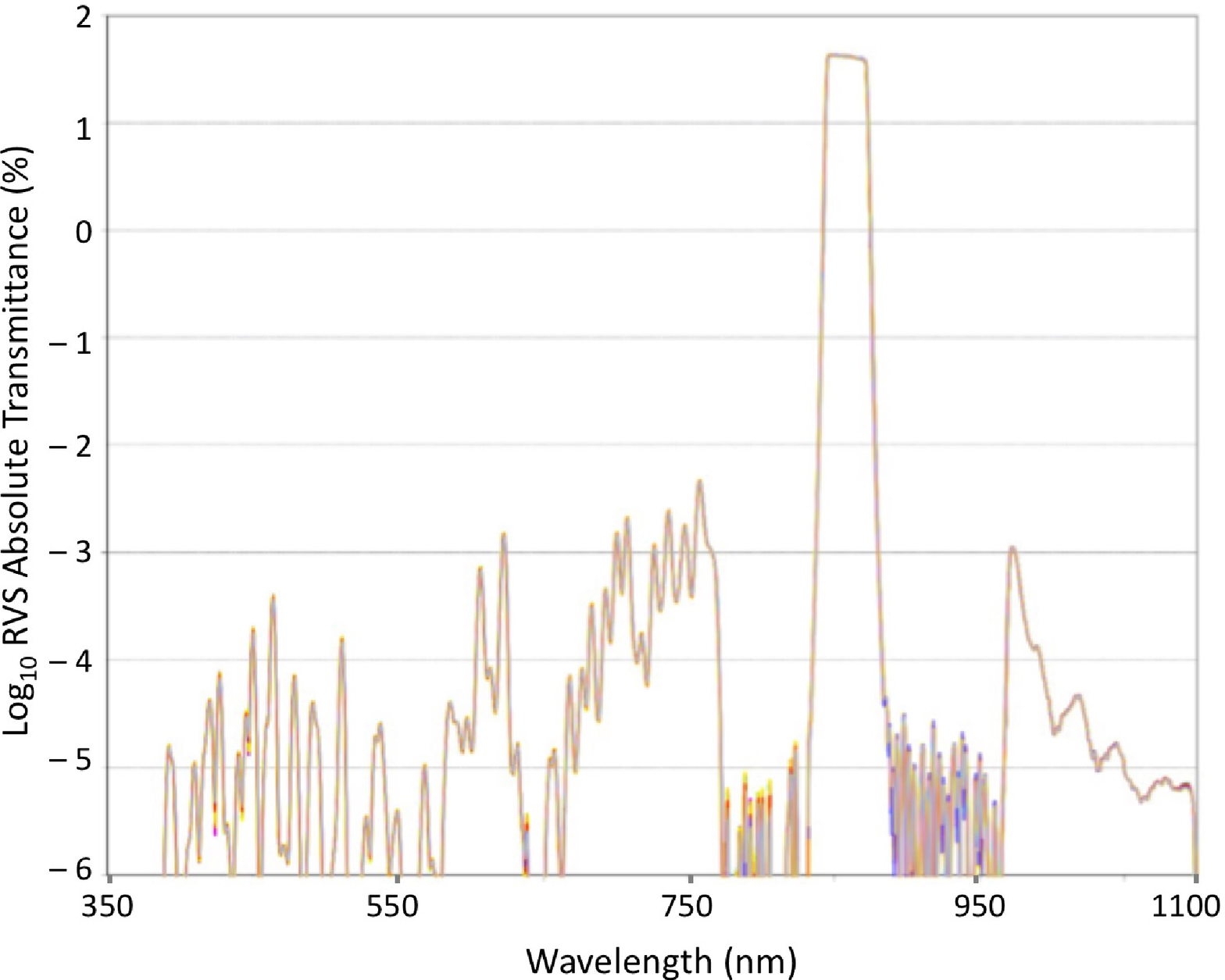}
\end{center}
\caption{RVS flight model bandpass showing the absolute transmittance with the specified bandpass in vertical red lines \textit {(top)} for the full optical chain and detectors, and on a log scale, the out-of-band rejection \textit {(bottom)}. The different colours distinguish between the different field points for both telescopes, and hence the slightly different angles at which the rays traverse the filter coatings. From \cite{Astrium:13}.}
\label{fig:bandpass}
\end{figure}

Table~\ref{tab:throughput} identifies the predicted overall pre-launch contributions at CDR to the RVS throughput, including the full optical chain and detectors. Owing to the excellent transmission of both the bandpass filter and the grating, this ranged between 0.40 and 0.47. The slope in the response results mainly from the decreasing detector quantum efficiency from 76\% at 847nm to 65\% at 874nm. The instrument was predicted to detect $1$ e$^{-}$ pix$^{-1}$ at $V=15.1$ giving a zero point (for which 1 e$^-$s$^{-1}$ is detected in the full bandpass) at $V=21.3$. Figure~\ref{fig:bandpass} shows the slightly lower throughput measured for the flight model. 

During flight model manufacture, it was found that spatial mid-frequency errors in the multilayers of the bandpass filter caused unacceptable performance degradation, and the filter was remade. However, for this element, the bandpass cutoffs were $\sim 2$ nm blueward, also resulting in a slightly narrower-than-specified instrument bandpass. This resulted in the astrophysically important \ion{Mg}{i} spectral line at 873.6 nm falling on the edge of the bandpass, degrading the accuracy with which it could be measured.

The wavelength-integrated rejection levels outside of the bandpass are required to be $\leq10$\%, and this is met for longer wavelengths, but marginally exceeded at shorter wavelengths at some field points, largely because of leakage at wavelengths between $700-750$ nm. 

\begin{table}[h!]
\begin{center}
\caption{\label{tab:throughput} Bandpass-averaged optical transmission, followed by the total photon detection fraction, including both optics and CCDs, for wavelengths at the extremes of the RVS bandpass. From \cite{Astrium:11}.}
\begin{tabular}{lccc}
\hline\\[-2mm]
\bf{Optical}        &  Per element  & Surfaces   &  Transmission  \\
Mirrors                 & 0.97 & 6 & 0.83  \\
Bandpass filter         &  &  & 0.95 \\
Grating                 &  &  & 0.80 \\
Prisms \& lenses & 0.98 & 8 & 0.85 \\
Contamination   &  &  & 0.93 \\
\underline{Microroughness} &  &  & \underline{0.99} \\
Total optical   &   & & 0.62 \\ [2mm]
\bf{Wavelength} & Optical &CCD & \bf{TOTAL} \\
847 nm & 0.62 & 0.76 & \bf{0.47} \\
874 nm & 0.62 & 0.65 & \bf{0.40} \\
\hline
\end{tabular}
\end{center}
\end{table}

The pre-launch colour conversion between the standard Johnson-Cousins photometric $V$ band and $G_{\rm RVS}$ is given in Tab.~\ref{tab:Grvs} \citep{Jordi:14}. This table also lists the conversions for the spectral types used in setting the performance requirements in Tab.~\ref{tab:requirements}.

\begin{table}[h!]
\caption{\label{tab:Grvs}
Pre-launch conversion between $V$ band and $G_{\rm RVS}$ \citep{Jordi:14} and for the spectral types in Tab.~\ref{tab:requirements} used by \cite{Astrium:11}. }
\begin{center}
\begin{tabular}{lll}
\hline\\[-2mm]
 \multicolumn{2}{l}{$G_{\rm RVS}$ = $V - 0.06 - 1.10 (V-I)$}   \\
B1V: & $G_{\rm RVS} = V + 0.18$  \\
G2V: & $G_{\rm RVS} = V - 0.87$ \\
K1III\,MP: & $G_{\rm RVS} = V - 1.22$ \\
\hline
\end{tabular}
\end{center}
\end{table}

\subsection{Spectral resolving power}

\begin{figure}[h!]
\begin{center}
\includegraphics[width=0.95\columnwidth] {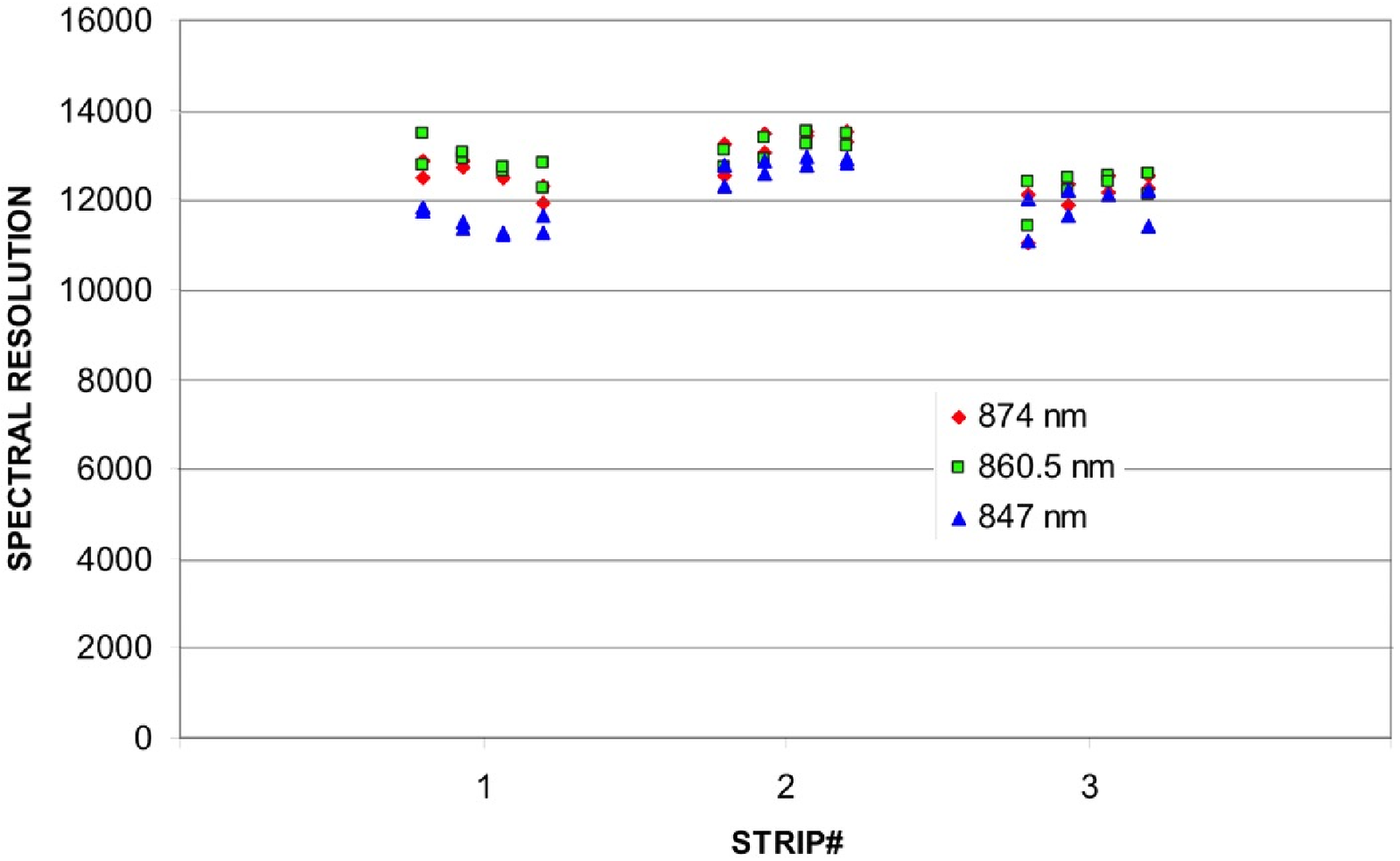}
\hspace*{5mm}\includegraphics[width=0.91\columnwidth] {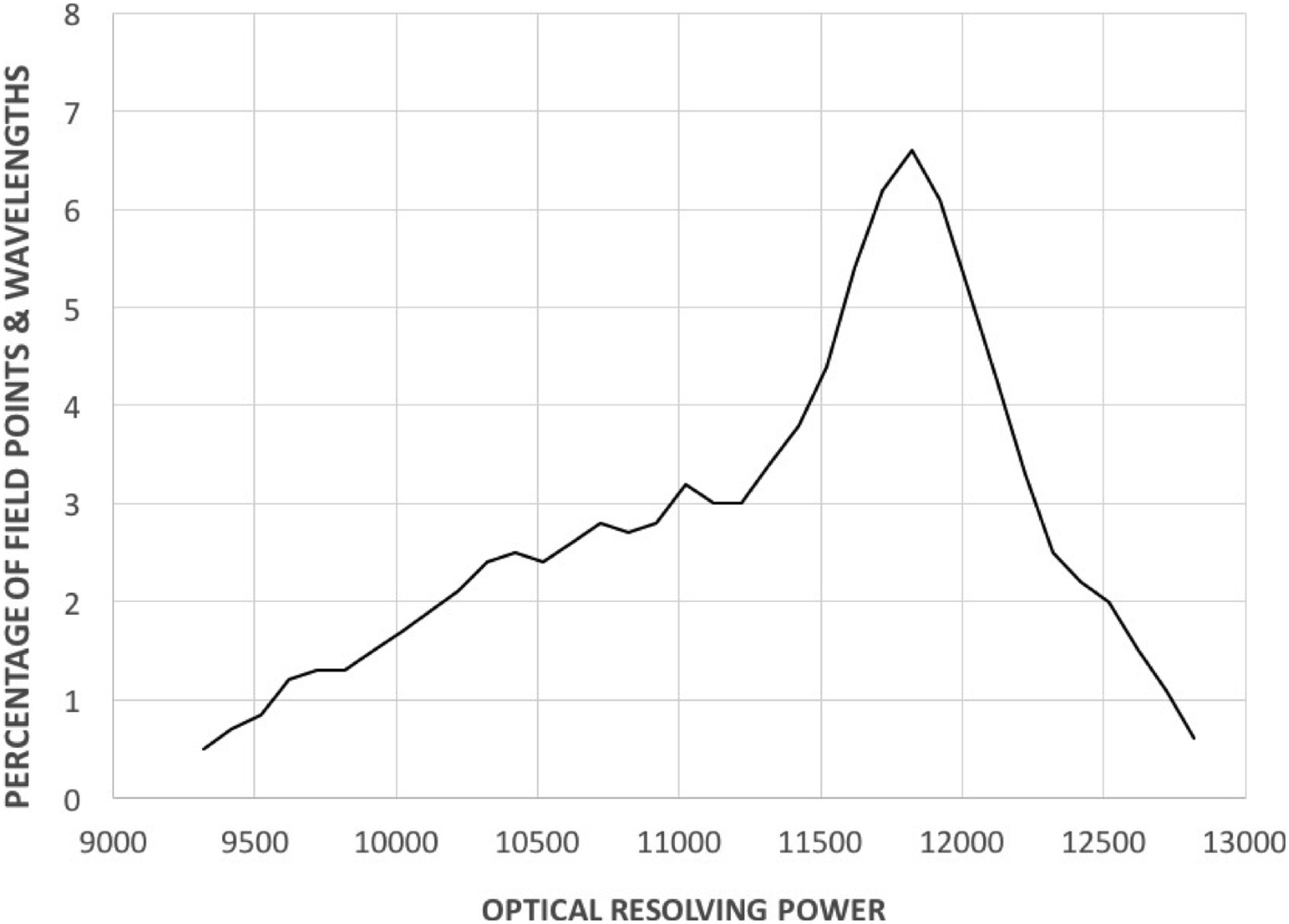}
\end{center}
\caption{\textit {(Top)} Average spectral resolving power for wavelengths at the extremes and centre of the RVS bandpass for each of the 4 CCDs in each strip and for each telescope. \textit {(Bottom)} Distribution of optical resolving power calculated over a grid of 13\,000 field points and wavelengths. From \cite{Astrium:11}. }
\label{fig:resolving_power}
\end{figure}

Figure~\ref{fig:resolving_power} shows the average spectral resolving power per CCD for the bandpass centre value and extremes, as well as the distribution of resolving power. This is  compliant with the permitted maximum and minimum levels in Tab.~\ref{tab:requirements}: 9\% of points are at a resolving power of $<10\,000$. The median is 11\,500. While average spectral resolution values for some CCDs lie above the $10\,500-12\,500$ range, the average is also compliant in total.
 
The dispersion law can be modelled by a quadratic polynomial in wavelength \citep{Astrium:08}, with the linear term being $\sim0.0245$ nm pixel$^{-1}$. The quadratic term is small, and there is a slight spatial variation over the RVS focal plane.

\subsection{Distortion}

Because \Gaia\/ operates in a scanning mode, optical distortion reduces the spectral resolving power and broadens the spatial profile. The average distortion (for both telescopes) displaces the light rays as they traverse the CCDs by 0.27 pixels along scan and 0.11 pixels across scan. 

\subsection{Noise performance}

The principle contributors to the noise in RVS spectra (in addition
to the Poisson noise of the spectra themselves) as expected pre-launch were the CCD readout noise, the external cosmic background, and the leakage from the BAM lasers that operate in the RVS band. An average readout noise of $3.7$e$^-$ for the full detection chain was measured at the payload module thermal vacuum testing, the most significant noise source when considered against the Poisson noise of the expected background flux of $0.6$ e$^-$ and BAM laser leakage of $0.3$ e$^-$. 

\subsection{Bias non-uniformity}

During the flight model testing programme, it was found that the electronic bias levels (the electronic signal corresponding to zero optical flux) in the PEMs were not constant, but varied in response to perturbations in the readout pattern in the serial register. One of the perturbations is the transition between the rapid flushing of unwanted pixels at MHz rates, and the slower reading of desired pixels at kHz rates; another is the pause in reading out the serial register at the time at which the parallel phases of the CCD are clocked for the TDI operation.
There are four of these per pixel and hence four `glitches' per serial readout. With these perturbations, the bias level drops sharply before recovering. 

The effect on \Gaia\/ as a whole is discussed fully in \cite{Hambly:18}, but it is discussed briefly here because of its particular impact for the RVS. It arises for two reasons: firstly in order to achieve a fine digitisation of the signal of $\sim$0.55 e$^{-}$, the overall gain of the detection chain is unusually high, a factor $\sim$6 higher than elsewhere in the payload; and, secondly, excursions of $\sim$70 e$^{-}$ (in the case of the flushes) can exceed the typical signal levels by two orders of magnitude. \cite{AllendePrieto:09a} analysed the impact of the bias non-uniformity on the radial velocity performance, and also on the fidelity of recovery of astrophysical parameters from the spectra. They found that the increase in radial velocity error, even after the application of a simple correction proposed by \cite{Astrium:09a} in which the boundaries of macrosamples were aligned (because the flushing patterns remain constant within a macrosample) was in excess of 10\% for stars at the faint limit. The impact of the spectral line distortions on the derivation of astrophysical parameters was more significant, with variations dependent on the spectral type and radial velocity of the star. 

\begin{figure}[h!]
\begin{center}
\includegraphics[width=0.7\columnwidth] {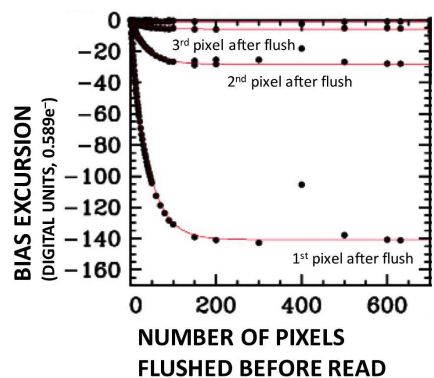}
\end{center}
\caption{Example of the calibration (red lines) of the bias non-uniformity (black dots) after a flush. Three groups of datapoints are evident: the lowest, corresponding to the maximum bias excursion, is the first pixel after the transition from flush to read; the middle and upper are the second and third pixels, respectively. The excursion depends strongly on how many flushes precede it (the horizontal axis), with the effect saturating after $\sim 200$ flushed pixels. The upwards excursion above the calibration line after 400 pixels is, in this particular case, the glitch caused by the pause in the serial readout.}
\label{fig:flush_calibration}
\end{figure}

While steps were taken to improve the stability of the hardware, options were limited on programmatic grounds. The adopted approach was to calibrate the effect, and an extensive campaign of laboratory measurements was executed. Figure~\ref{fig:flush_calibration} shows an example of the excursion after a flush, and it includes the effect of a glitch. It is evident that the time constant for recovery from the flushes is $1-2$ serial clock periods. While the glitches occur always at the same points for each TDI advance, flushes occur randomly relative to these, depending on the placement of the source windows and VOs. Analytic models were found to be effective predictors of behaviour, with the distribution of the residuals after correction at the $\sim$4 -- 5 e$^-$ FWHM level. These required parameters to be specific to each CCD-PEM pair and operating mode (HR, LR). The stability of the parameters in time and amplitude were not established in this campaign, which remained a concern until launch.

\subsection{Radiation damage}
\label{subsec:radiation_damage}

\begin{figure}[h!]
\begin{center}
\includegraphics[width=0.95\columnwidth] {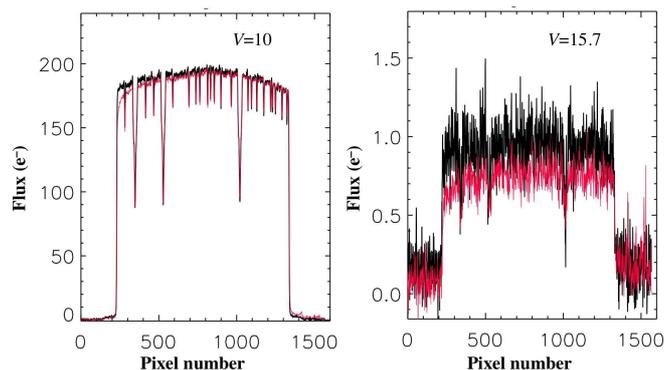}
\end{center}
\caption{Radiation damage effects to spectra as measured from laboratory testing using a spectral mask approximating a G2V star for flux levels corresponding to magnitude $V=10$ \textit {(left)} and $V=15.7$ \textit {(right)}. Spectra shown in black are the reference with no radiation damage, and those shown in red are the spectra after a radiation fluence of $10^9$ p$^+$ cm$^{-2}$ (10 MeV equivalent energy), the calculated fluence at end-of-mission for a launch date of 2012 December \citep{Astrium:11}. Noise levels are not equivalent in these plots, as a different number of individual spectra are combined in the two cases.}
\label{fig:radiation_damaged_spectrum}
\end{figure}

\begin{figure}[h!]
\begin{center}
\includegraphics[width=0.65\columnwidth] {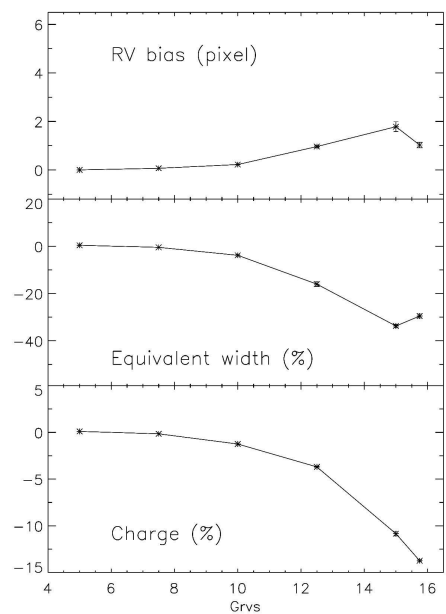}
\end{center}
\caption{Effect of radiation damage on RVS outputs as a function of magnitude ($G_{\rm RVS} \simeq V-0.87$ for G2V stars) for a fluence of $10^9$ p$^+$ cm$^{-2}$ (10 MeV equivalent). \textit {(Top)} Trapping and delayed release biases the shape of the lines, so that the velocity cross-correlation can be affected by ~1.5 pix ($\sim 13$ km s$^{-1}$) for the faintest stars. \textit {(Centre)} Spectral lines are filled in by the released traps, reducing the equivalent width by $\sim30$\%, while \textit {(bottom)} a total charge fraction exceeding 10\% is lost from the spectrum.}
\label{fig:radiation_damage_effects}
\end{figure}

Ions in the solar wind and from sources in the wider Universe
that impact on the detectors cause displacement damage, or "traps", in the Si lattice. These can impede the transfer of electrons from pixel to pixel in both the image area (parallel) and the readout register (serial). As an electron trapped in a damage site is released only after some time, its position on the sky will be recorded incorrectly, the scanning and readout having proceeded. This radiation damage has significant effects for the RVS: electrons in traps with long release times may be released only after the spectrum has entirely passed the trapping pixel, reducing the total counts in the spectrum; traps with intermediate release times modify the spectral energy distribution, in particular by removing the flux at the leading edge of the spectrum; and the traps with shorter release times modify the spectral line shapes and reduce the line amplitudes and equivalent widths \citep{AllendePrieto:09b}. The consequence is shown in Fig.~\ref{fig:radiation_damaged_spectrum}, where it is also evident that the damage effects are relatively greater for fainter flux levels.  The predicted impact on the radial velocity calibration, the equivalent width, and the charge loss is shown in Fig.~\ref{fig:radiation_damage_effects}. 

A substantial test and characterisation programme was carried out under the auspices of the \Gaia\/ Radiation Calibration Working Group and the \Gaia\/ DPAC in order to achieve an understanding of the radiation-induced effects. Models were developed to counter it, especially the charge distortion model described in \cite{Short:13}, which was sufficiently computationally fast to be used in forward-modelling approaches. These investigations demonstrated conclusively that
for the expected levels of radiation damage, single photoelectrons would survive the 4500 line transfers to reach the readout register. On the other hand, the low backgrounds expected in \Gaia, and especially in the RVS, were found to create particular susceptibility. Low-intensity diffuse optical background sources to flood the RVS focal plane were considered, but the consequent photon Poisson noise outweighed the positive effects of the improved charge transfer. Additionally, although the CCDs and PEMs were configured to permit an electrical injection of charge into the image-area pixels that the TDI would sweep through the image area, filling traps, given the frequency of these by comparison with the length of the RVS spectra, they were not expected to be used except in specific diagnostic circumstances. In the event, the higher scattered light backgrounds encountered post launch (Sec.~\ref{sec:post-launch}) provide some amelioration of this aspect of the radiation susceptibility.

\begin{figure}[h!]
\begin{center}
\includegraphics[width=0.85\columnwidth] {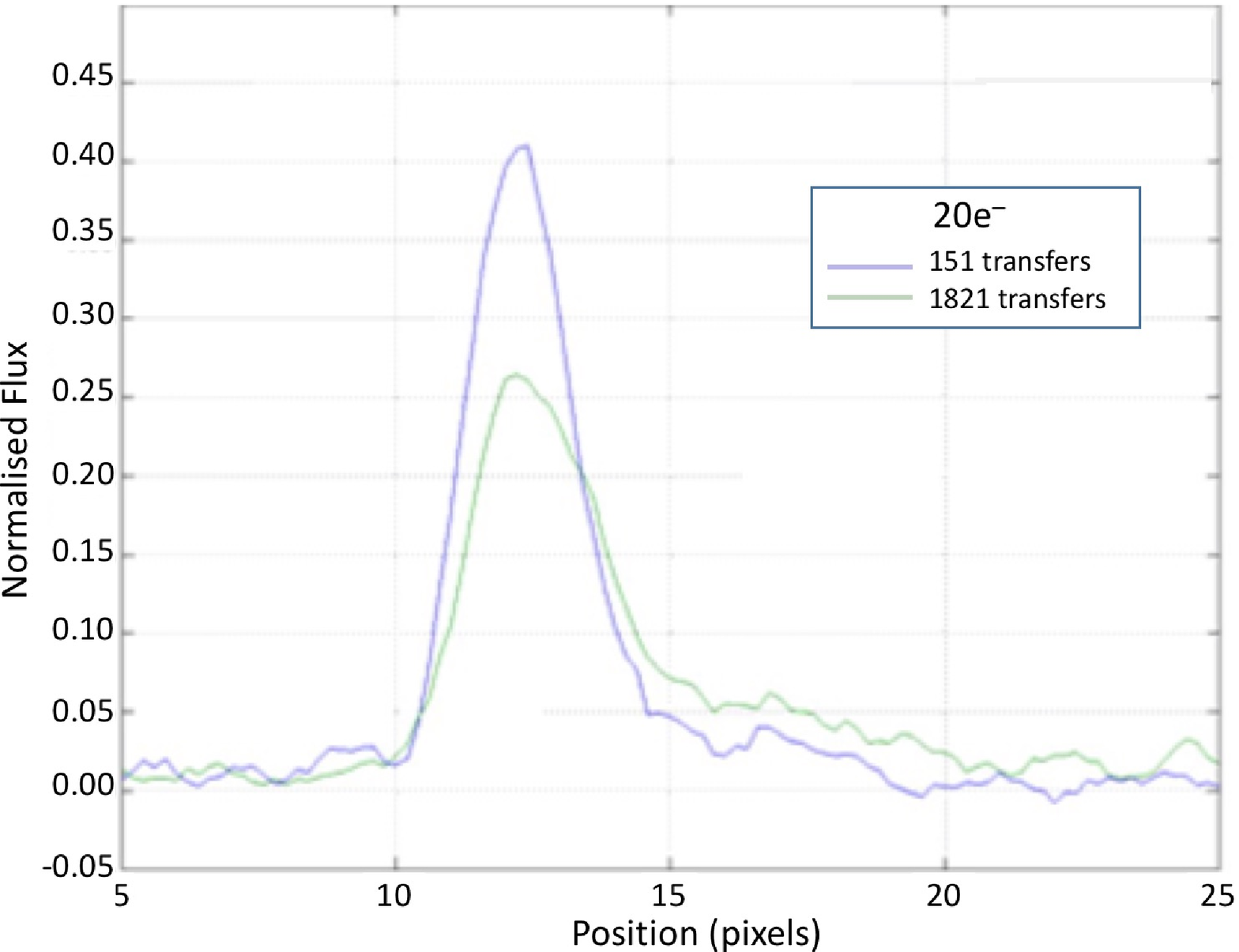}\\[2mm]
\includegraphics[width=0.85\columnwidth] {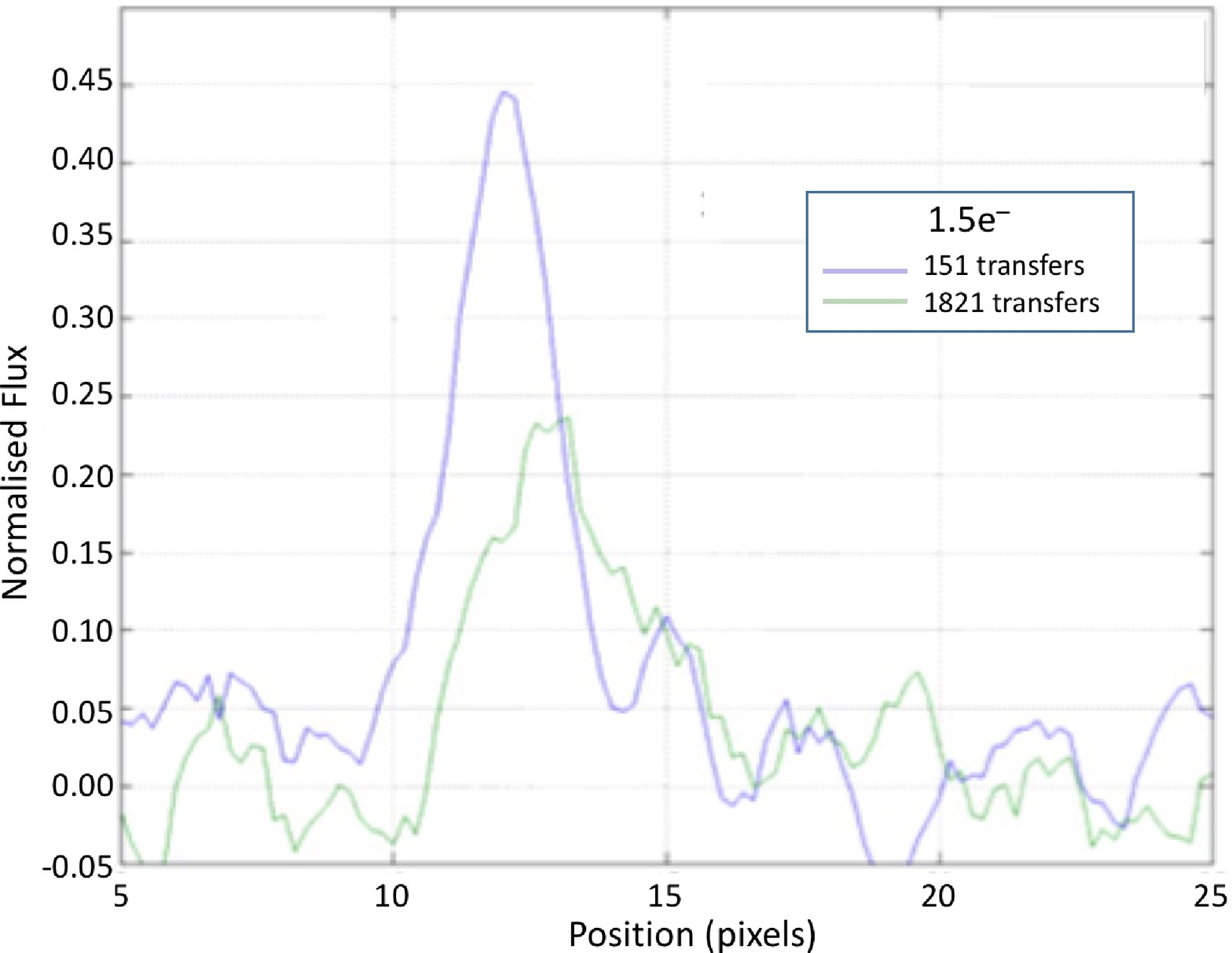}
\end{center}
\caption{Red variant CCD91-72 across-scan line-spread function for sources with 20 e$^-$  \textit {(top)} and 1.5 e$^-$  \textit {(bottom)} per HR sample (i.e.{\it } integrated over the line profile) corresponding to $V=11.9$ and $14.7,$ respectively, as a function of number of serial register transfers. This profile assumes the nominal optical performance, but no across-scan broadening from distortion or the forced precession in the scan law. From \cite{Astrium:09b}. }
\label{fig:serial}
\end{figure}

The effects in the readout register (serial) are complex, partly because the charge is transferred both at MHz rates (flushing unwanted data) and then at kHz rates (for reading), and hence is subject to traps with release times on both timescales. The MHz-rate traps dominate. The across-scan line spread function width increases with magnitude (so that it is worse for fainter stars) and also increases with the number of transfers (the distance from the readout node). In ground-testing, the red variant CCD91-72 was found to have more intrinsic traps (i.e.{\it }  already present at manufacture and not induced by radiation damage) than the other \Gaia\/ CCD variants. The effect of these serial register traps is shown in Fig.~\ref{fig:serial}, and it is significant for the flux levels expected in RVS spectra, with flux loss from the window and asymmetries in the profile that must be taken into account in overlapping windows (Sec.~\ref{subsec:to1D}).

\subsection{Effects of window collapse to one dimension}
\label{subsec:to1D}

The collapse of the two-dimensional to one-dimensional Class 1 and 2 windows at the detector level has little impact on spectra that are not overlapped by other spectra. Appropriate modelling of the spatial (across-scan) profile allows the fraction lost from the window to be corrected, and the calibration faint stars provide information on this profile down to the limiting magnitude of the instrument. However, overlapped spectra require special treatment in the ground data processing (called `deblending') to assign the flux detected in each (truncated) window correctly to each spectrum. The success with which this can be done  will depend on a number of factors, primarily the magnitudes of each star, their separation, the amount of spatial (across-scan) broadening resulting from the forced precession of the scanning law, and the optical distortion at that field point. In addition, radiation damage in the serial register changes the across-scan line spread function as a function of radiation fluence and of source magnitude (Sec.~\ref{subsec:radiation_damage}), so this will need to be taken into account as the mission progresses. 

\begin{figure}[h!]
\begin{center}
\includegraphics[width=0.75\columnwidth] {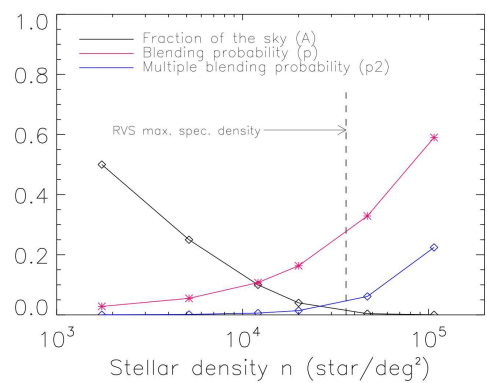}
\end{center}
\caption{Predicted frequency of overlapped and multiply overlapped spectra as a function of source density for each telescope. From \cite{AllendePrieto:08}.}
\label{fig:overlap_frequency}
\end{figure}

Pre-launch predictions of the fraction of spectral overlaps are shown in Fig.~\ref{fig:overlap_frequency}. This indicated that the average probability of overlap is 22\% (including both telescopes). The number of multiple overlaps is non-negligible, and this requires further treatment steps in the data processing. In each case, blended spectra will be more noisy after separation because of the imperfect deblending. However, because of the different path of the scan at each observation of a source, spectra that overlap at some epochs may not overlap at others, and it is possible to use the non-overlapped spectra and information from the photometer to assist in the deblending.

The lack of two-dimensional information within and around the window requires the faint source background to be modelled. Given the low flux levels, and because the sources are differently spatially arranged with respect to the selected source for each epoch of observation, this approach can be successful. However, in densely crowded regions, sources with $G_{\rm RVS}$ that would normally be assigned a window may not be, because the maximum number of assignable windows (72 per CCD) may be exceeded. These sources will be recorded as background. In this case, the background spectra may be moderately bright, and there may be no epochs at which a source spectrum is not overlapped by one source or another, so that a more specific treatment of the source data may be required.
 
\subsection{Dead time}
\label{subsec:dead_time}

The \Gaia\/ scan law \citep{Prusti:16} provides for more than the requisite average number of 40 transits across the RVS focal plane (Tab.~\ref{tab:requirements}). However, for a number of reasons, some transits may not be recorded. These include nominal orbital maintenance operations; inadequate resources for placement of the windows at the detection chain level in high-density regions; deletion in the onboard memory as a result of inadequate capacity, particularly when both telescopes are scanning the Galactic plane; and data transmission losses. All of these effects were modelled
by \cite{Astrium:11} to ensure that the average number of transits and radial velocity accuracies were met despite this dead time. 

The derived dead-time fractions are shown in Tab.~\ref{tab:RV_accuracy}. The dead time increases for fainter stars because of the $G_{\rm RVS}$ prioritisation assigned to windows and to the transmission of data. This creates a distribution (beyond that inherent from the scan law) in the number of transits recorded on the ground from star to star, and therefore in the expected end-of-mission radial velocity derived at each $G_{\rm RVS}$. This is particularly the case at the faint end, where the dead-time fraction exceeds 0.4.

\subsection{Radial velocity precision}

The radial velocity precision at CDR was predicted using a model taking into account the expected distribution of transits, dead time, spectral type, radiation damage effects, Poisson and readout noise, internal and external backgrounds, digitisation noise, across-scan collapse to one dimension for HR, and additionally, along-scan binning by 3 for LR. Spectra for each source were summed for all scans and cross-correlated with the same template as was
used to generate the spectrum. The computed value was increased by 20\% in accordance with \cite{deBruijne:05b}. The resulting performance is shown in Tab.~\ref{tab:RV_accuracy}.

\begin{table}[h!]
\begin{center}
\caption{\label{tab:RV_accuracy} Pre-launch predicted radial velocity precision for the specified stellar types in Tab.~\ref{tab:requirements} for end of mission. The precision does not include calibration residuals. Dead time combines the likelihood of all of the effects in Sec.~\ref{subsec:dead_time}. From \cite{Astrium:11}.}
\begin{tabular}{lcccc}
\hline\\[-2mm]
\bf{Spectrum}        &  Magnitude   & Dead &  Mode  & RV precision \\
                              &   $V$ band & time   &    & (km s$^{-1}$) \\
\hline \\[-2mm]
B1V             & 7      & 0.14 & HR  & 0.6  \\
G2V             & 13    &  0.34 & LR & 0.6 \\
K1III\,MP       & 13.5 &  0.34 & LR & 0.6 \\
\hline \\[-2mm]
B1V             & 12      & 0.34 & LR  & 8.5  \\
G2V             & 16.5    &  0.42 & LR & 12.8 \\
K1III\,MP       & 17      &  0.42 & LR & 13.3 \\
\hline
\end{tabular}
\end{center}
\end{table}

The radial velocity precision requirements from Tab.~\ref{tab:requirements} were required to be $\leq1$ and $\leq15$ km s$^{-1}$  for the brighter and fainter stars, respectively, in the upper and lower halves of Tab.~\ref{tab:RV_accuracy}. These predictions indicated that the key RVS radial velocity performance should be met with margin. It should be noted, however, that these predictions did not include the significant effects encountered in orbit, as discussed in the next section.

\section{Post-launch developments}
\label{sec:post-launch}

On 2013 December 19,  \Gaia\/ was launched on an accurate trajectory to its orbit insertion at L2. During this initial phase, the sunshield deployed and communications were established, and the payload underwent a controlled cool-down to its operational temperature (in the case of the RVS focal plane, 163K) during the transfer to L2. After spacecraft checkout, the payload was activated, and the first images were received on 2014 January 4. These trailed images from the non-spinning spacecraft indicated that the focus was approximately correct. The spacecraft spin rate was synchronised to the CCD readout rate as set by the onboard atomic clocks, and the focus of the two telescopes was optimised using the astrometric focal plane. The first RVS science data were received on 2014 January 17. Commissioning of the RVS then proceeded until the In-orbit Commissioning Review on 2014 July 18, using analyses for the initial data products tailored to the exploratory nature of these activities. 

\begin{figure}[h!]
\includegraphics[width=\columnwidth] {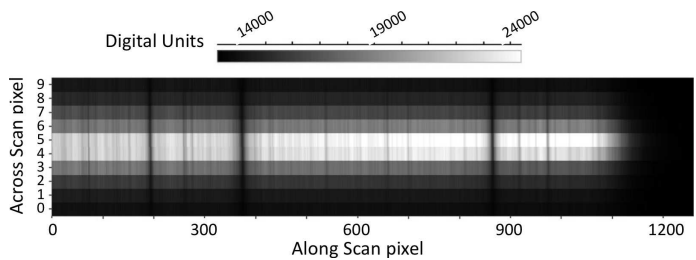}
\caption{Early RVS spectrum: Class 0 window ($12.6 \times 0.3$ mm on the CCD) of the $G_{\rm RVS}=6.2$ star with transit\_ID=4635432253571710\_1 from telescope 1 in FPA Strip 15, Row 4, observed on 2014 January 23. In this raw image, the bias level has not been subtracted and the aspect ratio of the window is widened in the across-scan direction for better visibility. The grey-scale is linear in digital units ($\sim 0.55$ e$^-$).}
\label{fig:class_0}
\end{figure}

\begin{figure}[h!]
\includegraphics[width=\columnwidth] {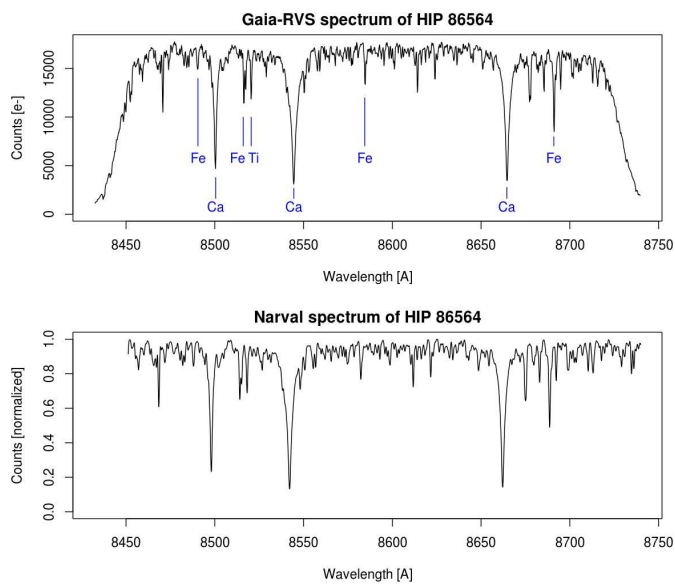}
\caption{\textit {(Top)} The first public spectrum from RVS, of the $V=6.67$ K5 star HIP 86564 identifying the major spectral features. The electronic bias has been subtracted and a wavelength calibration applied. This single CCD exposure has a signal-to-noise ratio $\sim125$. \textit {(Bottom)} A spectrum in the RVS spectral range of the same star taken with the NARVAL spectrograph at Observatoire Pic du Midi, convolved to the same spectral resolving power as the RVS spectrum. }
\label{fig:RVS_Narval}
\end{figure}

Figure~\ref{fig:class_0} shows one of the first RVS spectra. This is a Class 0 window preserving two-dimensional information. The focus is good even prior to optimisation, as is evident from the sharp absorption lines. In this window the spectrum is well centred in the 10-pixel width, with a small spectral tilt, but an adjustment in the spectral (along-scan) direction is required. Figure~\ref{fig:RVS_Narval} shows the first public\footnote{https://www.cosmos.esa.int/web/gaia/iow\_20140605} spectrum from RVS, of the $V=6.67$ K5 star HIP 86564. Lines of Fe and Ti are visible, and most prominently, the Ca triplet is clearly evident, with the instrument spectral bandpass centred on the Ca triplet. The quality of the data in this single 4.4s exposure is striking. Figure~\ref{fig:RVS_Narval} also shows ground-based data with a high signal-to-noise ratio for comparison, and it is evident that even spectral features with low equivalent width in the two spectra are in common.

\begin{figure*}[h!]
\begin{center}
\includegraphics[width=1.7\columnwidth] {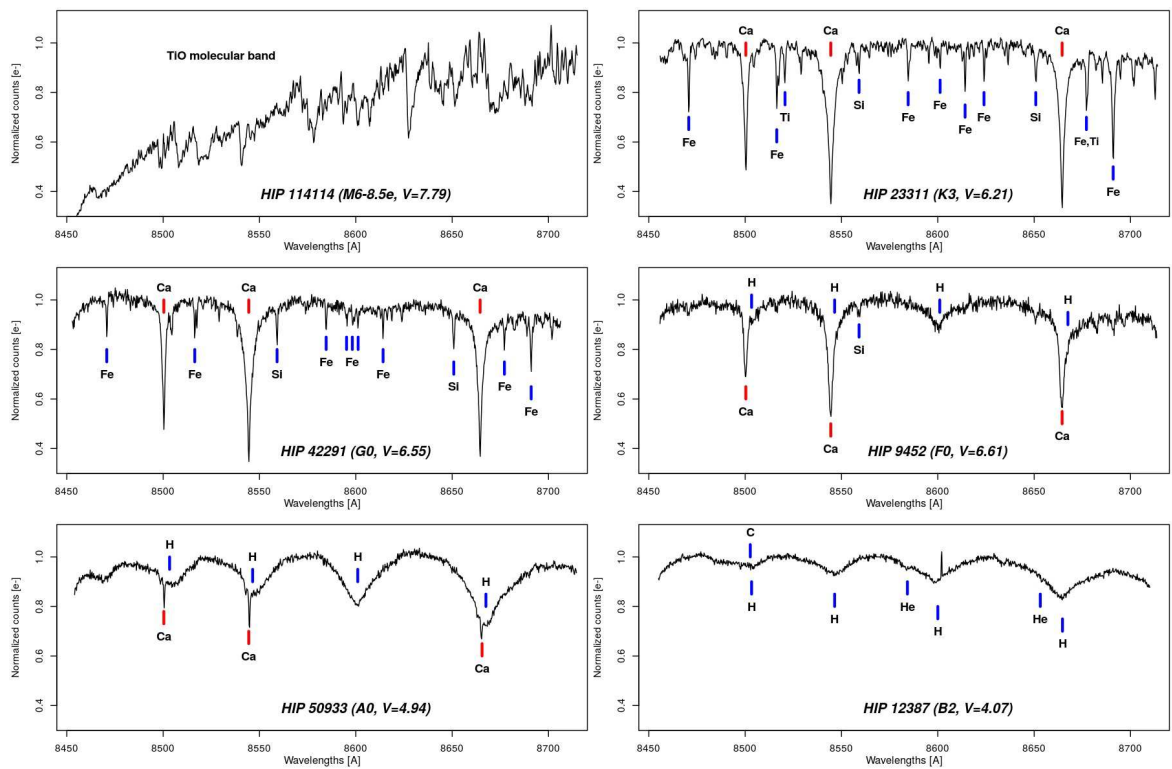}
\end{center}
\caption{RVS spectra of 6 \Hipparcos\/ stars from spectral type B2 to M6, with an identification of the major spectral lines in the RVS bandpass. In addition, although not evident in these spectra, diffuse interstellar bands are recorded at 862nm.}
\label{fig:spectral_types}
\end{figure*}

Figure~\ref{fig:spectral_types} shows sample RVS spectra\footnote{https://www.cosmos.esa.int/web/gaia/iow\_20141124} for spectral types B2 to M6. The dominant lines in hot stars are the Paschen series, with the Ca triplet dominating in spectral types F -- K, and TiO molecular bands for M stars. The richness of astrophysical diagnostics in this short wavelength range is
evident, vindicating the recommendations of the RVS Working Group (Sec.~\ref{sec:early}).

Broadly speaking, the in-orbit RVS spectra showed the characteristics expected pre-launch \citep{Astrium:11, Astrium:14b}. Nevertheless, it had become clear during investigations of the impact on the
RVS of the scattered light from the BAM laser almost immediately after receipt of the first RVS science data that unexpected variations were evident in the flux levels in the (nominally empty) VO windows. 

\subsection{Scattered light}
\label{subsec:scattered}

\begin{figure}[h!]
\includegraphics[width=\columnwidth] {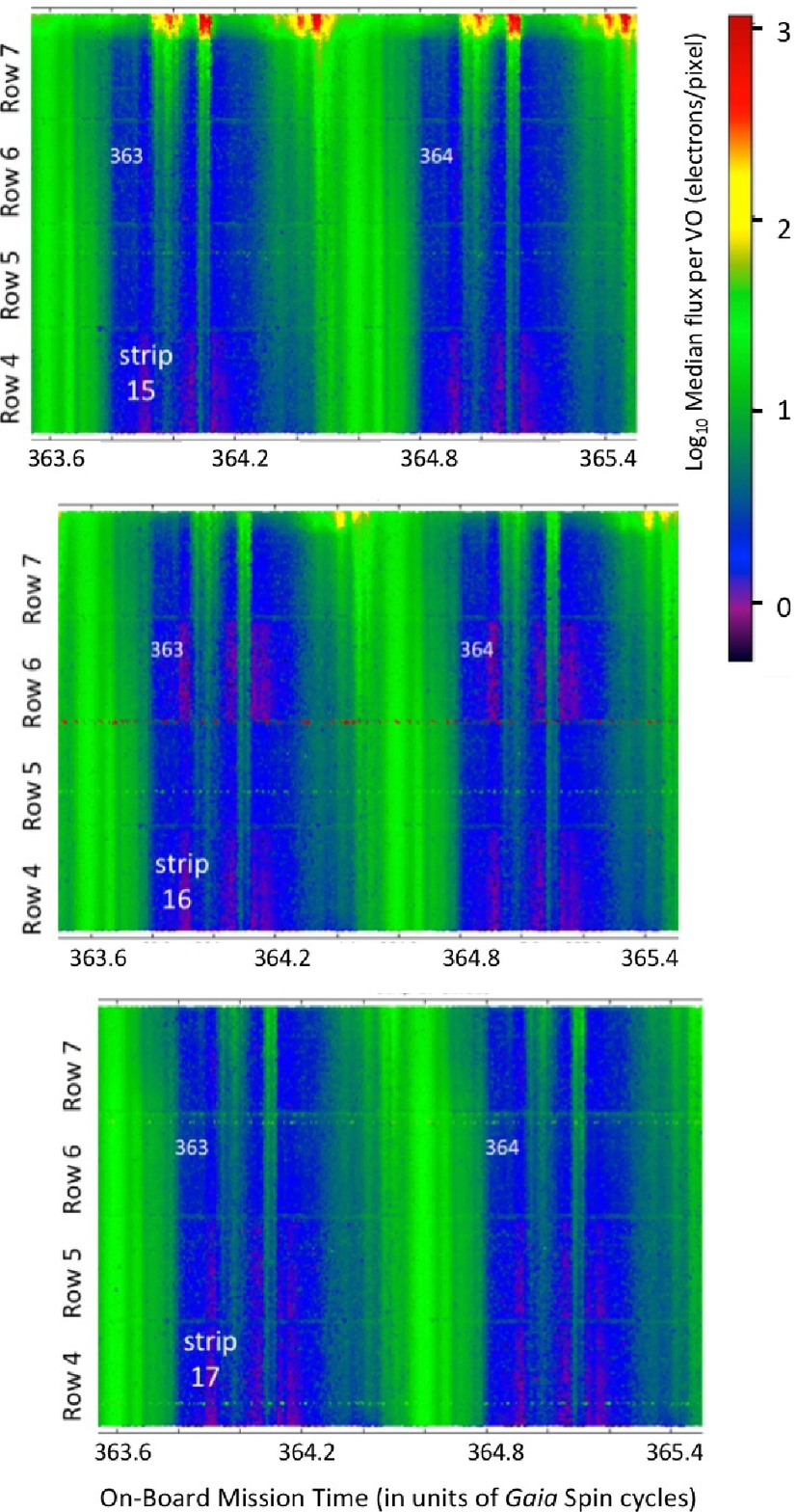}
\caption{Scattered light measurements early in the mission. The colour scale of the plots shows background level (for each VO, the median number of electrons per pixel per 4.4 s exposure) in logarithmic units as a function of onboard mission time in units of \Gaia\/ spin periods on the horizontal axis. The total duration is $\sim 12$ hr, covering \Gaia\/ spin periods 363 and 364 on 2014 January 30. The 4 rows of RVS CCDs are abutted, so that the plots show the across-scan dependence. There is a separate plot for each of the 3 CCD strips, each with an (exaggerated) displacement to denote their 4.4 s and 8.8 s later positions in the scanning.}
\label{fig:scattered_light}
\end{figure}

Figure~\ref{fig:scattered_light} shows the background light variation early in the mission. A pattern repeating on the \Gaia\/ 6hr spin period is evident. Further analysis based on the correlation of the flux levels with the satellite spin phase established that this variation was caused by scattered light from both the Sun (the broader features broadly common to all rows) and the bright stars/planets and Galactic Plane (sharper features most prominent in Row 7 at the top of the focal plane). The contribution from these sources changes relative to that of the Sun (and hence spin phase) as different sources are viewed and as the satellite spin axis precesses while the solar aspect angle is held constant at $45$\deg.

The origins of the increased scattered sunlight levels were traced to fibres at the perimeter of the flexible segments of the \Gaia\/ sunshield \citep{Astrium:14a,Prusti:16}, while those from bright stars/planets and the Galaxy disk arose from unexpected unbaffled optical paths in the payload module (Sec.~\ref{sec:baffles}). The scattered light background can be more than an order of magnitude higher than the expected $<0.8$ e$^{-}$ pixel$^{-1}$ per 4.4 s exposure, with solar scattered light values reaching $20-30$ e$^{-}$ pixel$^{-1}$ per exposure at some spin phases. At the top of Row 7, the scattered light level can exceed $10^{3}$ e$^{-}$ pixel$^{-1}$ per exposure, and may even reach detector saturation level. This development had significant implications for the RVS: as noted in Sec.~\ref{sec:requirements_lim-mag}, control of noise sources is critical in reaching the required radial velocity performance at lower flux levels where the instrument was to operate with expected background signals of $<1$ e$^-$ pixel$^{-1}$ per exposure. For a significant fraction of the spin period, the Poisson noise on the background now dominates the noise budget, exceeding that of the readout noise. In addition, exceptional care would be required for the background subtraction in order not to introduce biases into the velocity measurement for faint stars, and in brighter stars, for the measurement of line equivalent widths used for the determination of atmospheric parameters and individual abundances.

\subsection{Other new non-conformances}

The other significant non-conformances noted during the commissioning period such as the basic angle instability and the contamination buildup of water ice were less problematic for the RVS. The former affects the zero point of the wavelength scale, which is derived from the astrometric measurements, but at a low level compared to the end-of-mission systematic radial velocity error. The latter affects the RVS throughput, but less so than it affects the astrometric and photometric payload, owing to the operation of the RVS at far red wavelengths. The contamination is effectively removed by a de-contamination procedure \citep[during the commissioning, there were three of these; see][]{Prusti:16}. However, the thermal perturbations to the payload module as a result of the decontamination procedures require time to return to stability (adding to the dead time) and a refocussing, with consequent effects on the across-scan line-spread function and the spectral resolving power calibrations, as well as the throughput.

\section{Mitigations}
\label{sec:mitigations}

When the origin and nature of the contaminating scattered light (Sec.~\ref{subsec:scattered}) was understood, detailed simulations were carried out to evaluate the impact \citep{Katz:14b}, and the \Gaia\/ onboard software was modified to optimise the RVS operation in the new environmental conditions as follows:
\begin{enumerate}
\item the spectral sampling was optimised given the new noise balance;
\item the instrument limiting magnitude was reduced (brightened), taking into consideration the degraded signal-to-noise ratios;
\item the RVS windows were enlarged in order to measure the instantaneous straylight level. 
\end{enumerate}

To elaborate, the LR mode, i.e.{\it } the along-scan summing of the spectra by groups of 3~pixels (Sec.~\ref{subsection:window_scheme}), had been implemented to reduce the read-out noise by $\sqrt{3}$ as well as to minimise the telemetry budget by a factor 3. With the scattered light significantly larger than the readout noise, the utility of the on-chip summing to minimise the electronic noise is reduced. Moreover, the LR mode has drawbacks: the spectral resolution element is sampled with only one pixel;  it requires a separate and extensive set of calibrations; and the frequent switches between LR and HR modes appear to produce cross-talk with other CCDs. On 2014 July 10--17, shortly before the start of the nominal mission, the onboard software was modified to record all spectra in HR mode. 

Recording all (even faint) spectra in HR mode increased the RVS telemetry rate outside of the allocation, and compensatory measures were required. With the strong increase of the scattered light, the faintest spectra recorded near the instrument limiting magnitude contained almost no information, even when combined together at the end of the mission. To avoid using telemetry bandwidth for these spectra, the limiting magnitude was decreased from $G_{\rm RVS}\leq17$ to $G_{\rm RVS}\leq16.5$ on 2014 June 12, then to $G_{\rm RVS}\leq16.2$ on 2014 July 10, prior to the in-orbit commissioning review. This partly mitigated the telemetry increase resulting from the recording of all stars in HR mode. $G_{\rm RVS}=16.2$ was appropriate for the average scattered light level, but because of the large-amplitude fluctuations over the six-hour spin period of the satellite and the variations over the RVS focal plane (Fig.~\ref{fig:scattered_light}), a further optimisation took place in 2015 June when the limiting magnitude was adapted to the level of the instantaneous straylight in each VPU, varying from $G_{\rm RVS}=15.3 - 16.2$, following the straylight pattern.

Pre-launch, it was planned to measure the scattered light, then expected to arise from the laser in the BAM, using virtual object windows. While the BAM is relatively stable in time, allowing its scattered light contribution to be accumulated over long periods, the scattered light from the Sun and bright stars has strong variations over the six-hour spin period, long-term seasonal variations, peaks from bright stars, and strong local gradients at the top of CCD Row 7 (Fig.~\ref{fig:scattered_light}). The calibration of the post-launch scattered light therefore requires many more free parameters than expected pre-launch. An increase in frequency of background measurement was required, and in 2015 June, the RVS windows were therefore enlarged from 1260 to 1296 pixels by adding 3 pixels to each of the 12 RVS macro-samples in a spectrum. In a slitless spectrometer, the beginning and end of the window receive no source photons (as these wavelengths lie outside of the bandpass filter) but record the scattered light background.

Other measures were also considered, including a modification of the nominal width of the windows depending on the source magnitude and instantaneous background level; the following of spectral tilts (Sec.~\ref{subsec:spectral_tilt}) by adapting the window boundary half way along the spectrum; and an enhancement of the prioritisation of window overlaps. These enhancements were implemented onboard and commissioned, but were found not to enhance the performance significantly while at the same time introducing additional complexity, and therefore they are currently not used. 

In addition to the onboard software changes, modifications were made in the RVS data processing chain in order to calibrate and subtract the scattered light \citep{Sartoretti:18}.

\section{In-orbit characteristics and performance}
\label{sec:performance}

This section describes the in-orbit characteristics of the RVS, generally superseding the predictions made in earlier sections. Most of these characteristics are as they were known at the time of the In-Orbit Commissioning Review at the end of the commissioning (2014 July) and covering \Gaia\/ spin periods 308 -- 1048. However, included are relevant post-optimisation values (2015 April) that take into account the mitigations in Sec.~\ref{sec:mitigations} introduced to improve the performance in the presence of the higher-than-anticipated scattered light background. The longer-term instrument parameter trends as derived from calibrations within the data processing are
 available for Data Release 2 in \cite{Sartoretti:18} and \cite{Katz:18}.

Table~\ref{tab:characteristics} summarises the overall instrumental parameters. More details are provided in \cite{Panuzzo:15} and in the  \Gaia\/ Parameter Database.

\begin{table}[h!]
\setlength{\tabcolsep}{1mm}
\caption{\label{tab:characteristics} Summary of useful RVS parameters. }
\begin{tabular}{lll}
\hline\\[-2mm]
Exposure  & & 4.4167 s \\
\multicolumn{2}{l}{Typical number of transits} & 41, each over 3 CCDs \\[1mm]
Image scale & & 169.7 $\mu$m arcsec$^{-1}$ nominal  \\
Pixel scale & along scan   & 10 $\mu$m \\
                   &                      & 0.0589 arcsec nominal\\
                   & across scan & 30 $\mu$m \\
                   &                      & 0.1767 arcsec nominal \\[1mm]
Window size & along scan & 1260 pix until 2015 June \\
                  &                     & 12 macrosamples of 105 pix \\
                   &                     & 1296 pix after 2015 June \\
                  &                     & 12 macrosamples of 108 pix \\
                  & across scan & 10 pix except in overlaps \\ [1mm]
\multicolumn{2}{l}{Wavelength range}  & 845.0 -- 872.5 nm (FWHM)   \\
                  &                                          & 845.5 -- 872.0 nm (at 90\%)   \\ [1mm]
Mean dispersion & at 847nm & 0.0244 nm pix$^{-1}$ \\
                  &                 & 8.51 km s$^{-1}$ pix$^{-1}$ \\
                  & at 873nm & 0.0246 nm pix$^{-1}$ \\
                  &                 & 8.58 km s$^{-1}$ pix$^{-1}$ \\
\multicolumn{2}{r}{Telescope 1, Row 4 bottom} & 0.02440 nm pix$^{-1}$  \\
\multicolumn{2}{r}{Row 7 top}  & 0.02460 nm pix$^{-1}$ \\
\multicolumn{2}{r}{Telescope 2, Row 4 bottom}  & 0.02438 nm pix$^{-1}$ \\
\multicolumn{2}{r}{Row 7 top}  & 0.02461 nm pix$^{-1}$ \\[1mm]
\multicolumn{2}{l}{Digitisation}  & 0.539 -- 0.595 e$^-$ (dig. unit)$^{-1}$ \\
                  \hline
\end{tabular}
\end{table}

\subsection{Focus and spectral resolving power}

The best focus search was carried out in several stages with a final position for the commissioning identified during \Gaia\/ spin periods 662--682. The along-scan (spectral) resolving power and sampling averaged over each RVS CCD across the bandpass as measured over \Gaia\/ spin periods 680--697 (2014 April 16 -- 21) are given in Tab.~\ref{tab:resolution}. These were measured from a cross-correlation of Fe lines with a binary mask \citep{Panuzzo:15}. An alternative analysis using high-resolution ground-based spectra for comparison \citep{Katz:14a} resulted in an estimation of a slightly lower spectral resolving power ($\sim10\%$). Both these analyses yielded values that are lower than predicted in Fig.~\ref{fig:resolving_power}, but within specification. Alignment between RVS and the astrometric field is sufficiently accurate that requirements can be met with the optimal astrometric field focus; no compromise intermediate position is required. In HR mode the optical resolution is fully sampled at slightly more than Nyquist (the early spectra taken in LR during the commissioning phase are significantly undersampled). 

As noted in Sec.~\ref{sec:post-launch}, refocussing is carried out after each decontamination, so the values in Tab.~\ref{tab:resolution} change slightly each time this occurs.

\begin{table*}[h!]
\begin{center}
\caption{\label{tab:resolution} Mean spectral resolving power and number of pixels per optical resolution element for each detector and each field of view. From \cite{Panuzzo:15}.}
\begin{tabular}{llllll}
\hline
        &               & \multicolumn{2}{l}{Telescope 1}               & \multicolumn{2}{l}{Telescope 2}         \\
Row     &       Strip   &       Resolving power &       Resolution element (pix)   &       Resolving power &       Resolution element (pix)   \\ 
\hline \\[-3mm]
4       &       15      &       12\,587 & $     2.798 \pm 0.009 $ &       12\,065 & $     2.923 \pm 0.009         $ \\
4       &       16      &       12\,361 & $      2.849 \pm 0.009         $ &     12\,106 & $      2.912 \pm 0.008        $ \\
4       &       17      &       12\,240 & $      2.876 \pm 0.009        $ &       11\,954 & $      2.948 \pm 0.009        $ \\
5       &       15      &       12\,159 & $      2.891 \pm 0.008         $ &     11\,600 & $      3.032 \pm 0.010        $ \\
5       &       16      &       12\,430 & $      2.827 \pm 0.007         $ &     11\,809 & $      2.978 \pm 0.009        $ \\
5       &       17      &       12\,085 & $      2.908 \pm 0.008         $ &     11\,861 & $      2.965 \pm 0.009        $ \\
6       &       15      &       12\,021 & $      2.919 \pm 0.007        $ &       11\,523 & $      3.045 \pm 0.010        $ \\
6       &       16      &       12\,132 & $      2.891 \pm 0.007         $ &     11\,447 & $      3.066 \pm 0.011        $ \\
6       &       17      &       12\,148 & $      2.888 \pm 0.006         $ &     11\,901 & $      2.948 \pm 0.009        $ \\
7       &       15      &       12\,117 & $      2.890 \pm 0.007         $ &     11\,078 & $      3.160 \pm 0.010        $ \\
7       &       16      &       11\,885 & $      2.946 \pm 0.007        $ &       10\,983 & $      3.188 \pm 0.010        $ \\
7       &       17      &       11\,525 & $      3.038 \pm 0.008         $ &     11\,377 & $      3.077 \pm 0.009        $ \\
        \hline
\end{tabular}
\end{center}
\end{table*}

 The across-scan (spatial) line-spread functions arrived at during the final best-focus search average 3.5 and 2.8 pix FWHM \citep{Panuzzo:15} over the field of view for Telescope 1 and Telescope 2, respectively. During observations, this is broadened by both the transverse motion induced by the scanning law and the optical distortion in the field of view \citep{Panuzzo:14}. This causes the line-spread function width to vary from 2.25 -- 2.6 pix in the ecliptic scanning law (where the transverse motion is small) in the case of CCD Row 4 Strip 15 in Telescope 1. In the nominal scanning law, the range broadens to 2.25 -- 4.2 pix.
 
\subsection{Spectral tilt}
\label{subsec:spectral_tilt}

The RVS spectra are slightly tilted with respect to the CCD window boundaries. This contributes to flux loss from the ends of the spectra if they are not correctly centred in the window. The across-scan difference between the central position of the first and last macrosamples ranges from 3 -- 4 pixels, with the most positive tilts for Row 4 and the most negative for Row 7, and the change in tilt following a linear relationship. During \Gaia\/ spin periods 400 -- 437, there was an offset of 2 across-scan pixels between the spectra from the two telescopes. 

\subsection{Readout noise}

Commissioning-phase readout noise measured from the pre-scan pixels in the CCDs averaged 3.1 e$^-$. This was slightly better than that measured in the thermal vacuum testing, where the corresponding value was 3.7 e$^-$. DPAC measurements found slightly higher values (higher by $0.2 - 0.3$ e$^-$ ) \citep{Katz:14b}. The low noise levels vindicated the attention paid to minimising this noise source, which is still the dominant source during the low-background phases of the spin period. 

The readout noise for each CCD is given in Tab.~\ref{tab:readout_noise}.

\begin{table}[h!]
\begin{center}
\caption{\label{tab:readout_noise} Measured post-launch detector chain noise (CCD readout noise and PEM noise, including digitisation) in e$^{-}$ for the 12 RVS detectors in HR mode. From \cite{Astrium:14b}. }
\begin{tabular}{lcccc}
\hline
                &  Row 4   & Row 5   &  Row 6   &  Row 7 \\
\hline \\[-3mm]
Strip    15 & 3.1 & 3.1 & 3.0 & 3.0 \\
Strip    16 & 3.0 & 3.4 & 3.1 & 3.0 \\
Strip    17 & 2.9 & 3.4 & 3.1 & 3.2 \\
\hline
\end{tabular}
\end{center}
\end{table}

\subsection{Bias non-uniformity}

\begin{figure}[h!]
\begin{center}
\includegraphics[width=0.8\columnwidth] {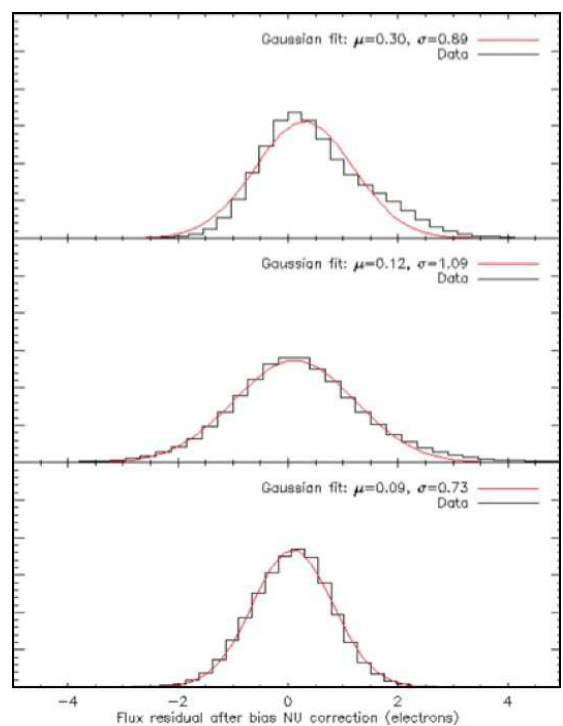}
\end{center}
\caption{Bias non-uniformity residuals for Strip 16, Rows 4 -- 6 for a set of special VOs taken on 2014 April 26. The calibration parameters derived from the VO set have been applied to the same set. Residuals in this case are at the level of $\sim 1$e$^-$ or less.}
\label{fig:BiasNU}
\end{figure}

The bias non-uniformity was calibrated from a set of  special VOs. When applied back to the set, residuals were in the range $0.52\leq\sigma\leq1.09$e$^-$ for all 12 detector chains, showing an excellent level of correction (Fig.~\ref{fig:BiasNU}). Several special VO patterns were taken over a period of weeks during commissioning to examine the stability of the effect;
this was the first time that this had been possible. During this time, the effect of gate activation was also examined; gates ensure that there are negligible numbers of photon events recorded in the VO. With the exception of one detector chain (Row 6, Strip 17), the coefficients were found to vary only slightly over the intervening period. When applied to routine VOs, the residuals were higher, at the $2-3$e$^{-}$ level; this and the existence of outlier points indicated that some improvement was required. 

\subsection{Radial velocity performance}

The key performance criteria for the RVS have from the outset been the radial velocity precisions in Tab.~\ref{tab:requirements}, and it is instructive to compare the in-flight performance to these requirements. The commissioning phase included a fortnight of performance verification (\Gaia\/ spin periods 772 -- 829). To maximise the number of repeated observations to achieve an assessment of the end-of-mission performance, the ecliptic pole scanning law was selected. 

\begin{figure}[h!]
\begin{flushright}
\includegraphics[width=0.97\columnwidth] {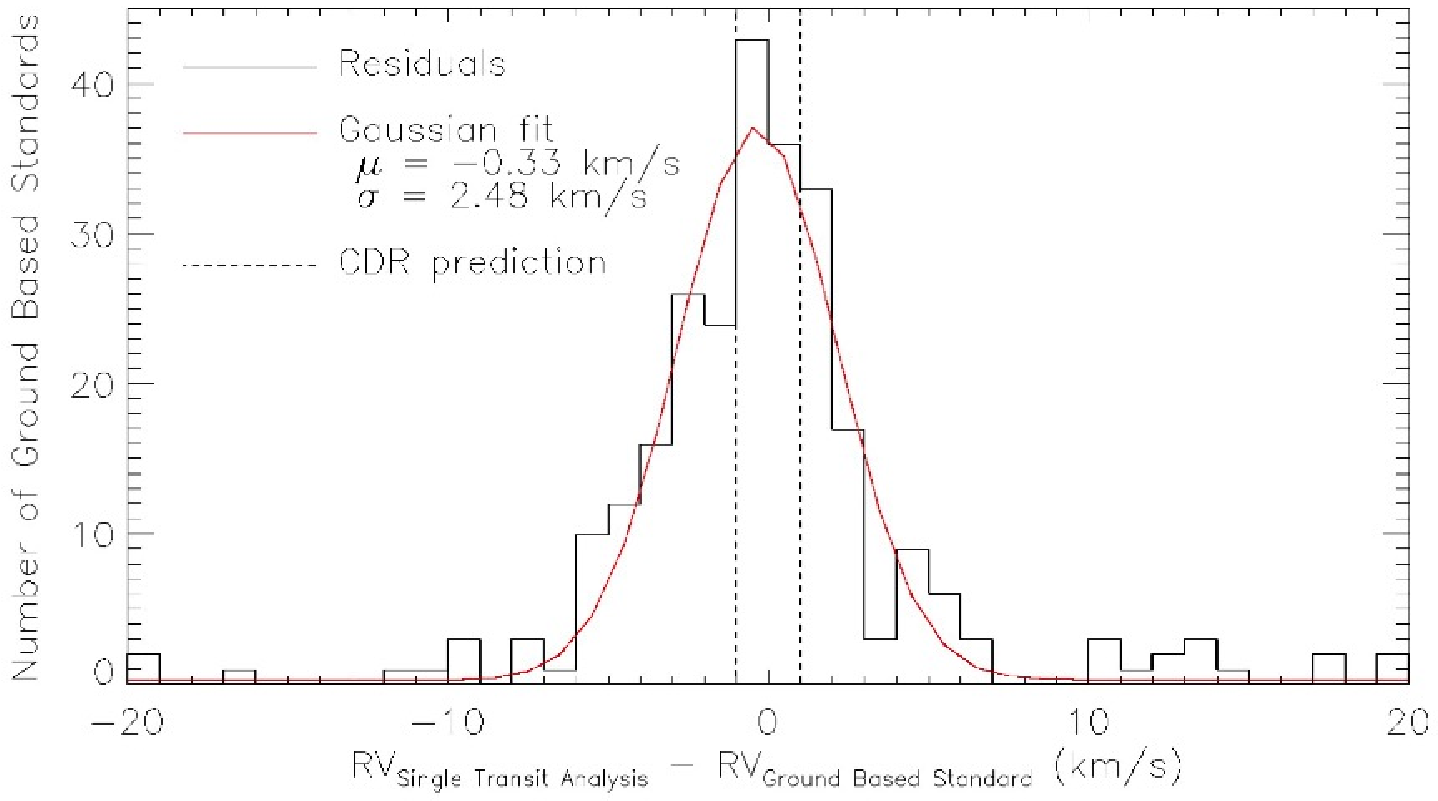}\\[4mm]
\includegraphics[width=\columnwidth] {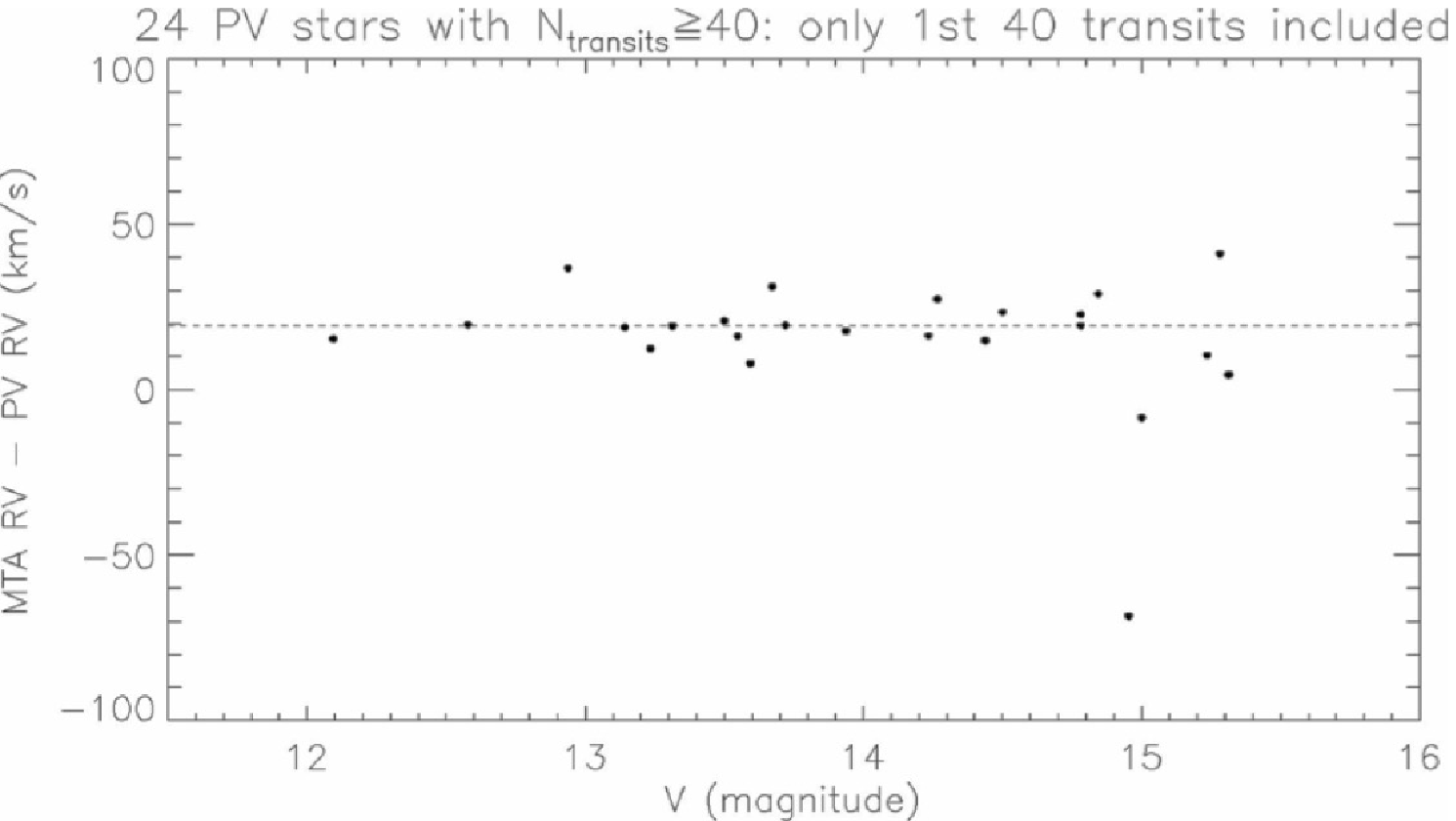}
\caption{\label{fig:RVs} \textit {(Top)} Radial velocity residuals by comparison with ground-based validation stars for single transits of the RVS focal plane (3 CCDs). \textit {(Bottom)} Radial velocity performance for stars with $\geq 40$ transits as a function of magnitude, compared with ground-based standards. These spectra were taken in LR mode using the ecliptic pole scanning law. The systematic offset of 19.6 km s$^{-1}$ is an artefact of a 2-pixel offset for the LR wavelength scale with respect to HR in the processing, and should be ignored. The cause of the large residual for one star of $V\sim15$ in this preliminary processing is unknown.}
\end{flushright}
\end{figure}

The upper panel of Fig.~\ref{fig:RVs} shows the distribution of residual radial velocities for single focal plane transits (three CCD strips) compared to ground-based standards \citep{Soubiran:18}. These stars were bright stars in the range $5 \leq G_{\rm RVS} \leq10$, taken in HR mode. The mean of the residuals is $-330$ m s$^{-1}$, almost consistent with the end-of-mission requirement (Tab.~\ref{tab:requirements}) of $300$ m s$^{-1} $ after only this limited period, and non-standard data processing. In addition, at this level, the mean is disturbed by some errors from the standards themselves. 

In respect of the radial velocity precision, the residuals in the upper plot of Fig.~\ref{fig:RVs} are for single transits rather than the average 41 expected at the end of mission. The dispersion in the fit of $2.48$ km s$^{-1}$ indicates that this should be easily met, but the distribution is dominated by stars brighter than that specified for $1$ km s$^{-1}$ in Tab.~\ref{tab:requirements} (for example $V>13$ in the case of a G2V star), so further analysis is required. See \cite{Katz:18} for a report in the \Gaia\/~DR2 dataset.

The spectra from stars with $\geq 40$ transits were combined and compared to fainter ground-based validation stars \citep{Fremat:17} to examine the radial velocity precision as a function of magnitude. This is shown preliminarily in the lower plot of Fig.~\ref{fig:RVs}. In this magnitude range during commissioning, only LR mode spectra were available. The number of available standards is small, but the results indicate that the radial velocity precision decreases after $V=15$ (the systematic offset is an artefact of the LR processing). 
A more detailed analysis of early mission data (Seabroke et al 2018, in prep) indicates, however, that for G to K stars in HR mode, the $15$ km s$^{-1}$ precision is reached at a $V$ limiting magnitude $15.8-16.5$, indicating (in the absence of significant radiation damage) a $\sim$0.5 magnitude shortfall of the original requirement (Tab.~\ref{tab:requirements}) in the 6000 -- 4500K temperature range because of the higher scattered light background.  This is 0.3 -- 0.6 magnitude better than the revised predictions\footnote{see https://www.cosmos.esa.int/web/gaia/science-performance} at the end of the commissioning, but at this early stage does not include any effects of in-orbit radiation damage.
 
\subsection{Radiation damage status}

Because of conservative assumptions and low solar activity levels during solar cycle 24 (\Gaia\/ was launched at the peak of this cycle), radiation damage to the \Gaia\/ CCDs has been well within the $10^9$ p$^+$ cm$^{-2}$ 10 MeV equivalent fluence that was designed for, with predicted end-of-mission values $\sim$10\% of this level \citep{Crowley:16}. Degradation in RVS performance as a result of radiation effects has been less than expected, aided also by the higher background levels from scattered light. In the short period from launch to the end of commissioning, no significant degradation was identified.

\section{Conclusion}
\label{sec:conclusion}

This paper has described the RVS on \Gaia\/, starting with the rationale for the inclusion of a spectroscopic instrument on a primarily astrometric mission. This has had the benefit of extending the mission from one that measures the dynamics in the Galaxy into a comprehensive facility for the wide-ranging investigation of the Galaxy structure and evolution. 

The RVS is not a typical spectrometer. Exposure times are set by the scanning requirements, resulting in extremely low signal-to-noise ratios in the spectra. This requires exceptional attention to the noise sources, driving all aspects of the design, from the throughput, selected bandpass, bandwidth, and spectral resolution to the noise performance and stability of the detection chain. Preservation of the information at the single photo-electron level and in the presence of radiation damage was required to be proven. The high data rates arising from relatively long spectra and short exposure durations required innovative and elaborate schemes to permit the information to be telemetered; these in turn had implications for the detection chain stability and noise. The important considerations driving the design of the RVS and its stages of development have been described here, together with the expected and in-orbit performance and mitigations taken to optimise this with the higher scattered light background both in the instrument and in the data reduction software. 

The data release policy\footnote{https://www.cosmos.esa.int/web/gaia/release\#} for \Gaia\/ RVS envisages the progressive release of increasingly fainter source data. This is because fainter sources require a sufficient number of transits to reach the signal-to-noise ratios in the accumulated spectrum necessary to achieve the specified radial velocity accuracy. It is also the case that an increasingly careful and elaborate approach is required to process data for stars at or near the limiting magnitude. Given the rapid increase in the Galactic distances probed by RVS measurements with increasing magnitude and the consequent rapid increase in the number of stars, the RVS scientific resource will be enhanced commensurably if the instrumental effects described here are calibrated and processed at a detailed level. In particular, attention is required on the bias non-uniformity, the scattered light and faint source background subtraction, correction for the radiation damage, the deblending, and the optimal combination of spectra (the ultimate fine corrections for the deblending and the radiation damage effects rely on {\textit {\textit{\textup{a priori}}}} knowledge of the radial velocity itself). From the extensive understanding of the instrument, gained both from the pre-launch analyses and tests from the in-orbit performance, it is clear what is required. The processing steps taken for the first RVS data release in \Gaia\/ Data Release 2 are described in \cite{Sartoretti:18} and \cite{Katz:18}, and further enhancements will be described with later releases.

The RVS is an exceptional resource for stellar and Galactic science. The scale of the survey is unparalleled and already exceeds by an order of magnitude the number of spectra recorded in previous surveys, with the expectation of more as the survey progresses. In terms of radial velocities, its scale and advantage is even greater by more than two orders of magnitude. While increased scattered light levels have reduced the precision of the radial velocities at fainter magnitudes, this is less than initially feared, and the mission extension will go some way to recover the initially expected radial velocity performance, especially if careful weighting of low background spectra
with higher signal-to-noise ratio is implemented in the data processing.

%

\begin{acknowledgements}

We  wish to acknowledge the role of Airbus Defence \& Space for their central role in the development of the \Gaia\/ RVS, and the \Gaia\/ Project Team at ESA
for their support of the RVS instrument within the \Gaia\/ payload.

This work has made use of results from the ESA space mission \Gaia\/, the data
from which were processed by the \Gaia\/ Data Processing and Analysis Consortium
(DPAC). 
Many of the authors are members of the DPAC, and their work has been supported by the following funding agencies: 
the United Kingdom Science and Technology Facilities Council and the United Kingdom Space Agency; the Belgian Federal Science Policy Office (BELSPO) through various Programme de D{\'e}veloppement
d'Exp{\'e}riences Scientifiques (PRODEX) grants; the French Centre National de la Recherche Scientifique (CNRS), the Centre National d'Etudes Spatiales (CNES), the L'Agence Nationale de la Recherche, the R{\'e}gion Aquitaine, the Universit{\'e} de Bordeaux, the Utinam Institute of the Universit{\'e} de Franche-Comt{\'e}, and the Institut des Sciences de l'Univers (INSU); the German Aerospace Agency (Deutsches Zentrum f{\"u}r Luft- und Raumfahrt e.V., DLR); the Italian Agenzia Spaziale Italiana (ASI) and the Italian Istituto Nazionale di Astrofisica (INAF); the Slovenian Research Agency (research core funding No. P1-0188); the Swiss State Secretariat for Education, Research, and Innovation through the ESA PRODEX programme, the Mesures d'Accompagnement, the Swiss Activit{\'e}s Nationales Compl{\'e}mentaires, and the Swiss National Science Foundation.

\end{acknowledgements}

\bibliographystyle{aa} 
\bibliography{refs} 

\end{document}